\documentstyle[10pt,epsfig]{article}
\begin{document}

\begin {center}
{\Large A study in depth of $f_0(1370)$}

\vskip 5mm
{D.V.~Bugg\footnote{email address: D.Bugg@rl.ac.uk}},   \\
{Queen Mary, University of London, London E1\,4NS, UK}
\end {center}

\begin{abstract}
Claims have been made that $f_0(1370)$ does not exist.
The five primary sets of data requiring its existence are
refitted with suitable Breit-Wigner amplitudes.
Major dispersive effects due to the opening of the $4\pi$
threshold are included for the first time; the $\sigma \to 4\pi$
amplitude plays a strong role.
Crystal Barrel data on $\bar pp \to 3\pi ^0$ at rest require
$f_0(1370)$ signals of at least 32 and 33 standard deviations
$(\sigma)$ in $^1S_0$ and $^3P_1$ annihilation respectively.
Furthermore, they agree within 5 MeV for mass and width.
Data on $\bar pp \to \eta \eta \pi ^0$ agree and require at least
a $19\sigma$ contribution.
This alone is sufficient to demonstrate the existence of $f_0(1370)$.
BES II data for $J/\Psi \to \phi \pi^+\pi ^-$ contain a visible
$f_0(1370)$ signal $>8\sigma$.
In all cases, a resonant phase variation is required.
Cern-Munich data for $\pi \pi$ elastic scattering are fitted well with
the inclusion of some mixing between $\sigma$, $f_0(1370)$ and
$f_0(1500)$.
Values of $\Gamma _{2\pi}$ for $f_2(1565)$, $\rho _3(1690)$,
$\rho _3(1990)$ and $f_4(2040)$ are determined.
\vspace{5mm}

\end{abstract}

\section {Introduction}
The $f_0(1370)$ plays a vital role in the spectroscopy of light
$J^P = 0^+$ mesons.
The role of experiment is to identify the resonances required by data
and determine their parameters: mass, width and branching ratios.
Many schemes exist for grouping the observed states into nonets.
The $0^+$ glueball is predicted at $\sim 1640$ MeV \cite {Glueball},
but with a sizable error from uncertainty in how to normalise the mass
scale.
The glueball would make an additional SU(3) singlet state,
mixing with neighbouring nonet states.
It is therefore necessary to scrutinise the evidence for each state.

Klempt \cite {KL1} and Klempt and Zaitsev \cite {KLno1370} question
whether there is an identifiable $0^+$ glueball and hence argue against
the existence of $f_0(1370)$.
Ochs has argued against it at conferences for two reasons.
He suggests that it can be confused with a broad `background' which
might obscure the analysis of the 1000-1500 MeV mass range
\cite {Ochs1}; secondly he argues that it has not been identified
definitively in Cern-Munich data \cite {Ochs2}.
The weakness of these arguments is that no attempt has been made to
refit the data where $f_0(1370)$ has been identified.
The main objective of the present work is to do just that.

The earliest evidence for the $f_0(1370)$ came from experiments on
$\pi \pi \to KK$ at the Argonne and Brookhaven laboratories in the
late 1970's.
A peak was observed in the S-wave at 1300 MeV in three sets of data:
Cohen et al.  \cite {Cohen}, Pawlicki et al. \cite {Pawlicki}
and Etkin et al \cite {Etkin}.
In those days it was called the $\epsilon (1300)$.
Further evidence appeared in the years 1992-6 from several experiments
in quick succession.
Amsler et al. reported a peak in the $\eta \eta $ S-wave at
$\sim 1400$ MeV in Crystal Barrel data on $\bar pp$ annihilation
at rest into $\eta \eta \pi ^0$ \cite {Amsler92} (as well as a further
peak at 1560 MeV, later identified as $f_0(1500)$).
Gaspero reported a $0^+$ state in $4\pi$ at $1386 \pm 30$ MeV in
$\bar p d$ bubble chamber data at rest in the reaction
$\bar pn \to 2\pi ^+3\pi ^-$ \cite {Gaspero};
the Obelix collaboration quickly confirmed this observation in
$\bar np \to 3\pi ^+2\pi ^-$ \cite {Obelix1}, quoting a mass of
$1345 \pm 12$ MeV.
Crystal Barrel reported a similar $0^+$ state in $4\pi$ at
$1374 \pm 38$ MeV in $\bar pp$ annihilation at rest to
$(\pi ^+\pi ^- \pi ^0 \pi ^0)\pi ^0$ \cite {Meyer}.

The earliest fits in which $f_0(1370)$ and $f_0(1500)$ appeared
together were made using early low statistics Crystal Barrel data
on $\bar pp \to 3\pi ^0$ at rest \cite {CBAR1}, \cite {CBAR2},
\cite {CBAR3}.
Publications using the full statistics data (used here) appeared in
1995 \cite {AmslerD} and 1996 \cite {FurtherA}.
The latter work involved a simultaneous fit \cite {CM1996} to
Cern-Munich data on $\pi \pi \to \pi \pi$ \cite {Hyams} and data on
$\pi \pi \to KK$.
The observation of nearby $f_0(1370)$ and $f_0(1500)$ states, both
dominantly non-strange, excited interest in the $0^+$ glueball.
Further extensive studies of $f_0(1370)$ and $f_0(1500)$ by
Crystal Barrel, Obelix and Omega collaborations may be traced via
the Particle Data Tables \cite {PDG}.

The present work fits simultaneously the 5 most definitive sets of
data available to me.
The first two are Crystal Barrel data on $\bar pp \to 3\pi ^0$
at rest in liquid hydrogen \cite {FurtherA} and gaseous
hydrogen. These two sets of data allow a clean separation of
annihilation from $^1S_0$ and $^3P_1$ $\bar pp$ states:
P-state annihilation is $\sim 13\%$ in liquid and $48\%$ in gas.
The combined fit also includes two sets of data for $\bar pp \to \eta
\eta \pi ^0$ in liquid and gas.
There are visible $f_0(1370)$ and $f_0(1500)$ peaks in both of these
sets of data.
The fifth definitive set of data comes from BES II for
$J/\Psi \to \phi \pi ^+\pi ^-$.
There is a visible $\pi \pi$ peak at 1350 MeV, attributed to
interference between $f_0(1370)$, $f_0(1500)$ and $f_2(1270)$
\cite {phipipi}.
Those data are refitted with and without inclusion of $f_0(1370)$.

Finally, an earlier analysis fitted data on $\pi
\pi \to KK$ and $\eta \eta$ \cite {Recon};
this analysis also fitted Kloe data on $\phi \to \pi ^0 \pi ^0 \gamma$
and $\eta \pi ^0 \gamma$.
The $\pi \pi \to KK$ data suggest the presence of $f_0(1370)$ but
cannot be considered definitive compared with the other sets of data.
This analysis remains consistent with parameters of $f_0(1370)$ found
here.

The Particle Data Group quotes  very large errors for mass and
width of $f_0(1370)$: $(1200-1500)-i(150-250)$ MeV.
From the present analysis, the error on the mass is very small:
$\pm 15$ MeV from systematics.
The resonance will still be referred to as $f_0(1370)$, despite
the fact that its peak position in the $\pi \pi$ channel comes out
nearly 100 MeV lower.
The Particle Data Group appears to be influenced by large variations
in masses and widths fitted to $4\pi$ data.
These variations are due to the fact that there has been no serious
attempt to include $\sigma \to 4\pi$ in these analyses.
A major objective here is to treat fully the dispersive effects due
to the opening of the $4\pi$ threshold, to which
$\sigma$, $f_0(1370)$ and $f_0(1500)$ all couple strongly.

The full form of the Breit-Wigner resonance formula,
\begin {equation}
f = 1/[M^2 - s - m(s) -iM\Gamma _{total}(s)]
\end {equation}
contains a real dispersive term $m(s)$ \cite {Nana}, which for the
$4\pi$ channel reads
\begin {equation}
m(s) =\frac {s - M^2}{\pi}\int \frac {ds'  M\Gamma
_{4\pi}(s')} {(s'-s)(s'-M^2)}.
\end {equation}
We shall find that $m(s)$ is large, indeed larger than $(M^2 - s)$ in
the Breit-Wigner denominator.
It turns out that this places severe limitations on the ratio
$\Gamma _{2\pi}/\Gamma _{4\pi}$ which can be fitted to data.
This point was not appreciated in earlier work.
Nevertheless, a good solution emerges naturally.

Here there is an interesting conclusion.
The fit includes explicitly the $s$-dependence of the $4\pi$ channel,
as far as present data allow.
However, the amplitude still produces an Argand diagram very close to
a circle and hence similar to a simple pole.
This justifies to some degree the common usage of a simple
Breit-Wigner formula in fitting data.

A third point concerns the coupling to $4\pi$ by the broad
component in the $\pi \pi$ S-wave related to the $\sigma$ pole; this
component will be called $\sigma$ as a short-hand. It plays an
essential role in fitting all data where a $\pi \pi$ pair is produced.
Let us review the situation briefly.

The $\pi \pi$ S-wave phase shift reaches $90^\circ$ at $\sim 900$ MeV.
There is an Adler zero in the elastic amplitude at $s \simeq
m^2_\pi/2$, just below the $\pi \pi$ threshold.
The resulting amplitude rises approximately linearly with $s$.
In production data from E791 \cite {E791} and BES II \cite {ompipi},
a strong peak is observed at $\sim 500$ MeV.
Both these production data and elastic scattering may be fitted
with the same Breit-Wigner denominator, but with an $s$-dependent
width and with different numerators for production and elastic
cases; for elastic scattering, the numerator contains the Adler
zero, whereas for the BES II production data fitted here it is
consistent with a constant.
The same variation of the phase shift with $s$ is observed in elastic
scattering and production data from 450 to 950 MeV within errors of
$\sim 3.5^\circ$ \cite {sigphase}.
The large displacement of the pole from 900 MeV arises from the
$s$-dependence of the width;
the Cauchy-Riemann relations control the $s$-dependence of the real
and imaginary parts of the amplitude as one extrapolates from the
physical $s$-axis.

The fit of Ref. \cite {Recon} to $\pi \pi \to KK$ and $\eta
\eta$ data and the Kloe branching ratio for $\phi \to \pi ^0 \pi ^0
\gamma$ determines ratios of coupling constants for the  $\sigma$
coupling to $KK$ and $\eta \eta$: $g^2(KK)/g^2_{\pi \pi} = 0.6 \pm 0.1$
and $g^2 _{\eta \eta}/g^2_{\pi\pi} = 0.2 \pm 0.05$.
The remaining unknown is the coupling of $\sigma$ to $4\pi$.
This will play an  essential role in the work reported here.

A vital question is whether this coupling to $4\pi$ eliminates
the requirement for the $f_0(1370)$.
Tornqvist  has suggested \cite {TornqvistA} \cite {TornqvistB} that a
second pole could appear in the $\sigma$ amplitude due to the opening
of the $4\pi$ threshold.
Could this explain $f_0(1370)$ as a non-$q\bar
q$ state? A related point is that Maiani et al. have suggested that
$f_0(1370)$ may be a molecular state \cite {Maiani}.

The layout of the paper is as follows.
Section 2 discusses the formalism and gives equations.
Amplitudes are expressed in terms of $T$-matrices for reasons
discussed there.
This raises some issues concerning how to fit elastic scattering.
Readers interested only in results may skip this section, but the issues
going into the formulae are outlined in words for the general
reader.
It is important to add that extensive fits to Crystal Barrel
and other data have been made by Anisovich and Sarantsev, using
K-matrix techniques \cite {AandS02}.
These analyses produce results closely similar to the present
work, and there is no conflict between the two analyses:
they should be regarded as complementary views.
The K-matrix analyses use several sets of data not available to me,
for example Crystal Barrel data on $\bar pp \to KK\pi$.
Their conclusion is that $f_0(1370)$ is needed, with mass and width
in close agreement to what is found here.
However, they do not address the question of how much the fit
changes if $f_0(1370)$ is omitted.

Section 3 is the heart of the paper, concerning fits to data with
and without $f_0(1370)$.
A suggestion made by Ochs is that the amplitude for the $\pi \pi$
S-wave should be fitted freely in magnitude and phase in bins of
$\pi \pi$ mass, without assuming a Breit-Wigner form.
Over the limited mass range 1100-1460 MeV, this is done and confirms
the assumption of a resonance form for the amplitude.

Section 4 describes the simultaneous fit made to Cern-Munich data on
$\pi  \pi $ elastic scattering.
They can be fitted slightly better with $f_0(1370)$ than without,
but cannot be considered definitive on this question.
Section 5 describes the fit to $\eta \eta \pi ^0$ data; there is a
visible peak in these data, sufficient alone to justify the existence
of $f_0(1370)$.

Section 6 describes the fit to BES data on $J/\Psi \to \phi \pi ^+\pi
^-$.
An important detail is that these data, together with data on $J/\Psi
\to \omega \pi \pi$ \cite {ompipi} and $\omega KK$ \cite {WKK} require
the existence of an $f_0(1790)$ distinct from $f_0(1710)$.
The $\phi \pi \pi$ data contain a clear $\pi \pi$ peak at 1790 MeV.
Data on $\phi \pi \pi$ and $\phi KK$ final states require a ratio
$BR[f_0(1790) \to \pi \pi ]/BR[f_0(1790) \to KK] > 3 $ \cite {phipipi}.
Data on $J/\Psi \to \omega KK$ contain a strong $KK$ peak due to
$f_0(1710)$; this peak is completely absent from $\omega \pi \pi$ data
which require $BR[f_0(1710) \to \pi \pi]/BR[f_0(1710) \to KK] < 0.11$
with 95\% confidence.
A single resonance must have the same branching ratios in all sets of
data, whereas these data differ in branching ratio by at least a
factor 22.
So two separate resonances are required.
Unfortunately, the PDG ignores these decisive results and continues to
lump $f_0(1790)$ with $f_0(1710)$.
In the present fits to $\bar pp \to 3\pi ^0$ data, $f_0(1790)$ is
included together with $f_0(1370)$ and $f_0(1500)$, though its eventual
contribution to the data is small.

Section 7 makes brief remarks on data for $\pi \pi \to KK$, and
section 8 remarks on the need for further analyses where the
opening of thresholds leads to large dispersive effects.
Section 9 makes concluding remarks.

\section {Discussion of Formulae}
There are two obvious ways of expressing amplitudes, using
$K$-matrices or $T$-matrices.
Each has some advantages, but also limitations.
It seems likely that neither is perfect, so approximations are needed
in either case.

Resonances appear directly as poles of the $T$-matrix.
The $\pi \pi$ S-wave is of primary concern.
For this amplitude, $K$-matrix poles are displaced strongly from
$T$-matrix poles.
For elastic scattering, $K \propto \tan \delta$ (where $\delta$ is
the phase shift), and $K$-matrix poles are at $\sim 700$ and 1200 MeV,
 whereas the $f_0(980)$ pole is at $998 - i17$ MeV.

In fitting production data, e.g. $\bar pp \to 3\pi ^0$, it is
obviously advantageous to use $T$-matrices, because resonances are
$T$-matrix poles.
The primary objective of the present work is to test whether the
$f_0(1370)$ is needed or not.
It is necessary to move its mass, width and couplings to all decay
channels in steps, so as to examine the effect on $\chi^2$.
This cannot be done readily using $K$-matrices, since the $f_0(1370)$ is
built out of a combination of $K$-matrix poles.
It is also necessary to remove $f_0(1370)$ from the fit and again
test $\chi^2$.
This cannot be done in a controlled way using
$K$-matrix poles: if one $K$-matrix pole is removed, all resonances are
affected.
For this reason, amplitudes will be written directly in terms of
$T$-matrices.

There is a second related point.
It is well known that minimisation routines converge best when
expressed in terms of eigenvectors, i.e. $T$-matrix poles.
Weak or questionable resonances appear as weak eigenvectors and can
be recognised immediately from the error matrix of the fit.

Yet another consideration is that earlier work fitting the
$\sigma$ pole and Cern-Munich data was done using $T$-matrices
\cite {sigpole}.
The formulae used there are readily expanded to incorporate the
$4\pi$ channel.
Also Ref. \cite {Recon} fitted data on $\pi \pi \to KK$ and $\eta \eta$
and also Kloe data with $T$-matrices; it is valuable to
maintain consistency with that analysis, for comparison of results.

\subsection {Elastic Scattering}
There are however questions about how to deal with elastic scattering.
Below the $KK$ threshold, the amplitude is confined to the unitary
circle.
Both $f_0(980)$ and $\sigma$ contribute, as do the low mass
tails of further resonances.
Experiment shows directly how to treat the overlap of these resonances.
In Cern-Munich data, the $\pi \pi$ phase shift rises dramatically
near 1 GeV from $\sim 90$ to $270^\circ$ due to the narrow $f_0(980)$.
The appropriate treatment below the $KK$ threshold is to add
phases, hence multiply $S$-matrices: $S = \exp (2i \delta)$.

Above the inelastic threshold, multiplying $S$-matrices gives a fit of
rather indifferent quality.
It is clear that other factors must be relevant.
If one solves a relativistic Schr\" odinger equation using a
trial potential which reproduces $\sigma$, $f_0(980)$ and $f_0(1370)$,
the solution is automatically unitary and analytic.
One finds that amplitudes differ from both
(i) the product of individual $T$-matrices for each resonance,
(ii) the sum of $K$-matrices.
Neither gives an accurate parametrisation.
The main problem appears to be that resonances mix via processes of
the form $<\sigma |\pi \pi |f_0>$ or other intermediate channels
$KK$, $\eta \eta$, etc.
Mixing is strong for elastic scattering, since the amplitudes
are at the unitary limit when one takes account of all channels.
The mixing gives rise to well known level-repulsion.
This repulsion is still highly significant one full-width away
from the resonance mass.

Formulae for mixing have been given by Anisovich, Anisovich and
Sarantsev \cite {AAS97}
and will be reproduced here in a slightly modified notation.
For the 2-resonance case, the denominator $D(s)$ of one resonance
may be written
\begin {equation}
D_{11}(s) = M_1^2 - s - m_1(s) - iM_1\Gamma_1^{tot}(s)
 - \frac {B_{12}(s)B_{21}(s)}{M_2^2 - s - m_2(s) -
 iM_2\Gamma_2^{tot}(s)}.
\end {equation}
 Mixing arises via $B_{12}$ which in general is complex and may be
 $s$-dependent.
 The propagator matrix describing two resonances is then
 \begin {eqnarray}
 \nonumber
 \hat {D} &=& \left| \begin {array}{cc} D_{11}&D_{12} \\
                             D_{21}&D_{22} \end {array}\right| \\
 &=& \frac {1}{(M_1^2 - s - m_1(s))(M_2^2 - s - m_2(s))}
 \left| \begin {array}{cc} M_2^2 - s - m_2(s) & B_{12} \\
                       B_{21} & M_1^2 - s - m_1(s) \end {array}\right|.
 \end {eqnarray}

 Because $\sigma$ overlaps strongly with $f_0(980)$, $f_0(1370)$ and
 $f_0(1500)$, this mixing has been included explicitly in fitting
 elastic data.
 It turns out that inclusion of mixing between these pairs of states
 leads to an excellent fit using constant values of $B_{12}$.

 It is also instructive to expand the denominator of the last term of
 eqn. (3) off resonance using the binomial theorem.
 The result is
 \begin {equation}
 D_{11}(s) = [M_1^2 - s - m_1(s) -iM_1\Gamma_1^{tot}]
  - \frac {B_{12}B_{21}}{M_2^2 - s}
    \left( 1 + \frac {m_2(s)+iM_2\Gamma_2^{tot}}{M_2^2  - s} \right).
 \end {equation}
 From the last term, one gets contributions of the same sign to
 $\rm {Re}\, D_{11}(s)$ and $\rm {Im}\, D_{11}(s)$, while from the
 first term $[M_1^2 - s - im_1 -iM_1\Gamma _1^{tot}]$ contributions
 have opposite signs.
 The result is to rotate the phase of the amplitude, which derives
 purely from $D(s)$.
 This rotation is large when resonances overlap strongly.
 Unless the mixing is included explicitly, one must expect that the
 resonance denominator may need to be multiplied by a phase factor
 $\exp (i\phi)$.
 This was indeed observed in Ref. \cite {Recon}, where it was
 sufficient to take $\phi$ as constant.
 This point will be relevant in fitting $1^{--}$ and
 $2^{++}$ states to elastic data.

 \subsection {The continuation of amplitudes below thresholds}
 Consider as an example $\pi \pi \to KK$.
 It is necessary to continue this amplitude below threshold;
 in $T$-matrix language this is the analytic continuation of the
 $T_{12}$ component.
 As a result, there are contributions to $\pi \pi$ elastic scattering
 from sub-threshold $\pi \pi \to KK \to \pi \pi$.
 However, caution is needed in making this continuation.
 The $\pi \pi \to KK$ amplitude is proportional to phase space
 $\rho _2 = \sqrt {1 - 4M^2_K/s}$ for the $KK$ channel.
 Below threshold, this continues analytically as $i\sqrt {4M^2_K/s -
 1}$.
 The analytically continued amplitude rises rather strongly below
 threshold.
 If one is not careful, this continued amplitude can make dominant
 contributions to elastic scattering below the inelastic threshold.
 This is counter-intuitive.
 A particular case arises for $f_2(1565)$, which couples strongly
 to $\omega \omega$.
 If one does nothing about the large sub-threshold $\omega \omega$
 contribution, it produces big interferences wth the nearby
 $f_2(1270)$ and can distort rather severely the mass and width
 fitted to $f_2(1270)$.

 The answer to this point is straightforward.
 Amplitudes above threshold contain form factors due to the finite
 radius of interaction forming the resonance.
 In Ref. \cite {Recon}, data on $\pi \pi \to KK$ were fitted
 empirically to an exponential form factor $\exp (-5.2k^2)$,
 where $k$ is momentum in the $KK$ rest frame in GeV/c;
 a study of $f_2(1565)$ in $\bar pp \to (\omega \omega )\pi ^0$ at
 rest in Crystal Barrel data also requires a form factor above the
 $\omega \omega$ threshold of similar strength \cite {Hodd}.
 The analytic continuation can be evaluated below threshold using
 a dispersion integral.
 This is not an accurate procedure because of uncertainties
 in the $\pi \pi \to KK$ amplitude above the available mass range.
 However, the qualitative feature emerges that a rapidly falling form
 factor is required below threshold as well as above.
 In Ref. \cite {Recon} this sub-threshold form factor was fitted to
 Kloe data with the result $\exp (-8.4|k|^2)$.
 This empirical cut-off will be adopted here for sub-threshold
 coupling to $KK$, $\eta \eta$ and $\omega \omega$.

\begin {figure}  [htb]
\begin {center}
\vskip -34mm
\epsfig{file=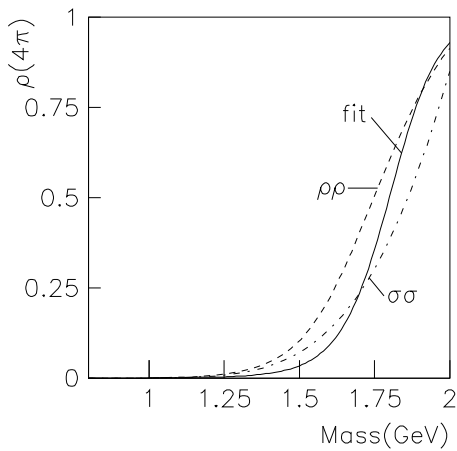,width=8cm}\
\vskip -10mm
\caption{$4\pi$ phase space for $\rho \rho$ (dashed), $\sigma \sigma$
(chain curve) and the fit adopted here (full curve)}
\end {center}
\end {figure}

 \subsection {Treatment of the $4\pi$ channel}
 The $4\pi$ phase space volume may be modelled \cite {CM1996}
 by the production of two resonances ( $\rho \rho$ or $\sigma \sigma$):
 \begin {equation}
 \rho_{4\pi}(s) = \int ^{(\sqrt {s} - 2m_\pi)^2}_{4m^2_\pi}
 \frac {ds_1}{\pi}\int ^{(\sqrt {s} - \sqrt {s_1})^2}_{4m^2_\pi}
 \frac {ds_2}{\pi}\frac {8|p||p_1||p_2|}{\sqrt{s s_1 s_2}}
 |T_1(s_1)|^2|T_2(s_2)|^2FF(s),
 \end {equation}
 where $p_1$ and $p_2$ are momenta of pions from decays of each
 resonance in its rest frame, and $p$ stands for the momenta of the
 $\rho$ or $\sigma$ in the centre of mass frame.
 In Ref. \cite {Nana}, extensive illustrations are shown of the
 dependence of $\rho _{4\pi }(s)$ on exponential form factors $FF$.
 These factors begin to play a significant role at $\sim 1.45$ GeV.
A form factor
 \begin {equation}
  FF = \exp [- (s - 1.45^2)]
 \end {equation}
 is chosen in present work with $s$ in GeV$^2$.
 If one were fitting data on $\pi \pi \to 4\pi$, this form factor
 would be rather important.
 However, for the present study of $f_0(1370)$ and $f_0(1500)\to
\pi \pi$ it
 has only rather small effects within errors.
 This is because
 the $\Gamma _{4\pi}$ term in the Breit-Wigner denominator
 cuts off the $\pi \pi$ channel strongly at high mass;
 the $f_0(1370) \to \pi \pi$ amplitude is already quite
 small at 1.45 GeV, where the form factor begins.
 The $f_0(1500)$ is sufficiently narrow that the effect on the
 line-shape from $4\pi$ inelasticity is rather small.
The one place where the form factor is important is in suppressing
high mass contributions to $m(s)$.

Fig. 1 shows $\rho \rho$ phase space as the dashed curve and
 $\sigma \sigma$ phase space as the chain curve.
 They are rather similar.
 Their relative contributions to each resonance are poorly known.
 The strategy here is to parametrise $\rho _{4\pi}$ empirically as
 \begin {equation}
 \rho _{4\pi} = \frac {\sqrt {1 - 16m^2_\pi /s}}
 {1 + \exp [\Lambda (s_0 - s)]}.
 \end {equation}
 The parameters $\Lambda $ and $s_0$ in the Fermi function of the
 denominator are optimised in the overall fit, with the result
 $\Lambda = 3.39$ GeV$^{-2}$, $s_0 = 3.238$ GeV$^2$.
The result is shown by the full curve on Fig. 1.

The dispersive contribution $m(s)$ to Breit-Wigner amplitudes is
evaluated numerically at 10 MeV steps of mass, and the programme
interpolates quadratically in mass using the nearest 3 bins.
The dispersion integral is the same for $\sigma$ and all $f_0$'s,
except for (i) a subtraction at the resonance mass M, where the
real part of the amplitude is zero, (ii) a scaling factor depending
on the coupling constant to $4\pi$.
The sub-routine for doing the principal-value integral is
available from the author if it is needed in other cases.

\subsection {Explicit equations for resonances}
Formulae for the $\sigma$ amplitude follow the same form as used in
earlier work on the $\sigma$ pole \cite {sigpole}, except for the
inclusion of $m(s)$ for $4\pi$.
Equations will be repeated here for completeness.
The elastic amplitude is written
\begin {equation}
T_{11}(s) = N(s)/D(s).
\end {equation}
The numerator contains an Adler zero at $s = s_A \simeq 0.41 m^2_\pi$.
For fits to BES data on $J/\Psi \to \omega \pi \pi$, the numerator is
taken as a constant.
Channels $\pi \pi$, $KK$, $\eta \eta$ and $4\pi$ will be labelled 1 to
4. The propagator of the $\sigma$ is given by
\begin {eqnarray}
D(s) &=& M^2 - s - g_1^2\frac {s - s_A}{M^2 - s_A}z_{s} - m(s)
-iM\Gamma_{tot}(s) \\
M\Gamma_1(s) &=& g^2_1\frac {s - s_A}{M^2 - s_A}\rho_1(s) \\
g_1^2 &=& M(b_1+b_2s)\exp [-(s - M^2)/A] \\
j_1(s) &=& \frac {1}{\pi}\left[2 + \rho _1  ln_e \left(
\frac {1 - \rho _1}{1 + \rho _1}\right) \right] \\
z_{s} &=& j_1(s) - j_1(M^2)  \\
M\Gamma_2(s) &=& 0.6g_1^2 FF^2_2(s) \\
M\Gamma_3(s) &=& 0.19g_1^2 FF^2_3(s) \\
FF_i(s) &=&\exp (-\alpha |k|_i^2) \\
M\Gamma_4(s) &=& Mg_4\rho _{4\pi }(s)/\rho _{4\pi}(M^2).
\end {eqnarray}
The value of $\alpha$ is 5.2 above thresholds and 8.4 below.
The numerical coefficient 0.19 in eqn. (16) has been revised
very slightly using the branching ratio fitted in
Section 5 between $\eta \eta$ and $\pi \pi$.
Values of $b_1$, $b_2$, $A$ and $M$ of eqn. (12) are given below
in table 6.

Resonance denominators for $f_0(1370)$, $f_0(1500)$ and $f_0(1790)$
are taken in the form of Eq. (1).
An important detail is that the factor $(s - s_A)/(M^2 - s_A)$ of
Eq. (10) is also used for $\Gamma_1$, $\Gamma _2$ and $\Gamma _3$ of
$f_0(980)$, $f_0(1370)$, $f_0(1500)$ and $f_0(1790)$.
That is, the Adler zero is included into the widths of all $0^+$
resonances.
Otherwise, parameters of $f_0(980)$ are taken initially from the
BES determination \cite {phipipi}, but are re-optimised within
the sum of statistical and systematic errors when fitting present data.
For $f_0(1500)$, the ratio $\Gamma _2/\Gamma_1$ is
taken from the PDG average.
For $f_0(1370)$, it is taken from Ref. \cite {AandS02}, where extensive
Crystal Barrel data on $\bar pp \to KK\pi$ are fitted.
However, in practice the ratio $\Gamma _2/\Gamma_1$ has rather
have little effect here;
only the full width of the resonance is crucial.
Values of $\Gamma_3/\Gamma _1$ are determined in Section 5, but again
have little effect.
The same ratios are used for $f_0(1790)$ as for $f_0(1500)$, in
the absence of good data for $f_0(1790)$ in the $KK$ channel; since the
$f_0(1790)$ signal is weak, this is of no consequence.
A trial has been made including into $f_0(1500)$ a weak coupling to
$\omega \omega$ with a coupling constant a third of that for known
decays to $\rho \rho$; the effect is negligible.

The $f_2(1270)$ is parametrised using
\begin {eqnarray}
\Gamma_{2\pi}(s) &=& \Gamma _{2\pi}(M^2)
\frac {k^2D_2(k^2)}{k^2_rD_{2}(k^2_r)}    \\
\Gamma_{4\pi}(s) &=&\Gamma _{4\pi}(M^2)\frac {\rho_{4\pi}(s)}
{\rho_{4\pi}(M^2)}.
\end {eqnarray}
Here $k$ is the pion momentum in the $f_2$ rest frame
and $k_r$ is the value on resonance.
The $D_2$ are Blatt-Weisskopf centrifugal barrier factors.
Expressions for them are given in Ref. \cite {CM1996} at
the end of Section 2.1; the barrier radius optimises at
$0.75 \pm 0.04$ fm.
The $KK$ and $\eta \eta$ channels are treated in the same way.
From the Particle Data book, $\Gamma _{2\pi}/\Gamma_{tot} = 0.847$ on
resonance.
This value is not sufficiently accurate for fitting
Cern-Munich data and is re-optimised to 0.802, since the relative
heights of $\rho(770)$ and $f_2(1270)$ are important.
This value may be accounting for a form factor
in $\pi p \to \pi \pi p$ over the mass range between these
two states.

The $\rho(770)$ is parametrised like $f_2(1270)$ using the $D_1$
centrifugal barrier factor.
Its mass and width optimise at 778 and 153 MeV, close to PDG values.
Trials were made including coupling to $KK$ and $\omega \pi$, but had no
significant effect on the fit to Cern-Munich data
compared with uncertainties in the $\rho(1450)$ contribution.

The $f_2(1565)$ presents a problem.
It is known from Crystal Barrel data \cite {Hodd} that it couples
strongly to $\omega \omega$.
It is expected to couple to $\rho \rho$ a factor $\sim 3$ more
strongly.
However, there are unfortunately no data to confirm this strong $\rho
\rho$ coupling. Consequently it is dangerous to fit $f_2(1565)$ with
the full $s$-dependence and $m(s)$ used for $f_0$'s. In view of the
fact the $f_0(1370)$ and $f_0(1500)$ line-shapes come out close to
those of simple poles, it is therefore fitted with a Breit-Wigner
amplitude of constant width, except that the $\omega \omega$ channel is
added explicitly. This channel has an important effect in cutting off
the line-shape above resonance, see Fig. 9 below.

Other $1^{--}$, $2^{++}$, $3^{--}$ and $4^{++}$ resonances are likewise
fitted with Breit-Wigner amplitudes of constant width, for simplicity
and ease of comparison with other work.
In fitting elastic data, each amplitude is multiplied by a fitted
factor $\exp (i\phi)$.
This is done rather than including mixing between all resonances, since
the number of mixing terms becomes too large, and furthermore the
resonances are poorly separated at present.

\begin {figure}  [htb]
\begin {center}
\vskip -28mm
\epsfig{file=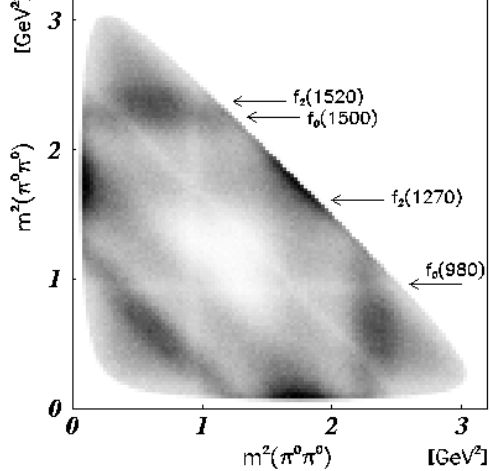,width=6.5cm}\
\vskip -5mm
\caption{The Dalitz plot for $\bar pp \to 3\pi ^0$ in liquid hydrogen}
\end {center}
\end {figure}

\begin {figure}  [h]
\begin {center}
\vskip -18mm
\epsfig{file=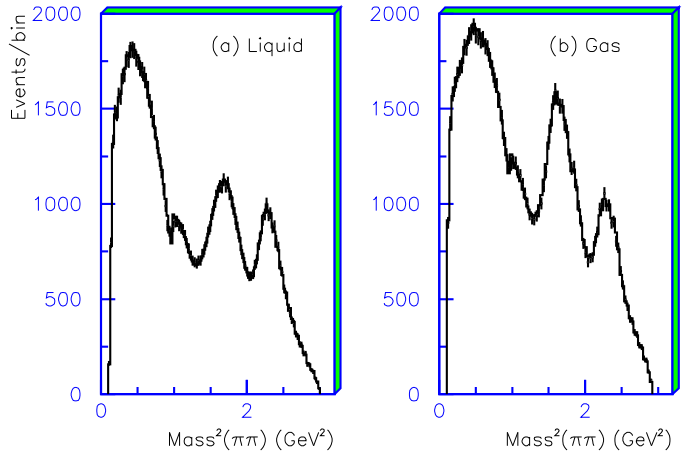,width=9.5cm}\
\vskip -6mm
\caption{Fits to the $\pi \pi$ mass proection in (a) liquid, (b) gas}
\end {center}
\end {figure}

\section {Fit to $\bar pp \to 3\pi^0$ at rest}
Fig. 2 shows the Dalitz plot for data in liquid
hydrogen at rest; it is similar for gaseous hydrogen.
[This figure is from Ref. \cite {AmslerD} where $f_2(1520) \equiv
f_2(1565)]$. Mass projections for both sets of data are shown  in Fig.
3. There are conspicuous peaks due to $f_2(1270)$ and $f_0(1500) +
f_2(1565)$. Narrow bands due to $f_0(980)$ are also visible.
Ingredients in the fit are $\sigma$, $f_0(980)$, $f_2(1270)$,
$f_2(1565)$, $f_0(1370)$, $f_0(1500)$ and a weak $f_0(1790)$.

P-state annihilation is best determined by data in gas, where it
makes up $48 \%$ of events, close to the predicted 50\% from
calculations of Stark mixing \cite {Stark}.
This is mostly annihilation with orbital angular momentum $L = 1$ in
the transition, but a minor detail is that there is a little
$L = 3$ production of $\pi ^0 f_2(1270)$ and $\pi ^0 f_2(1565)$.
The fit to liquid data uses $^3P_1$ and $^3P_2$ components scaled down
between gas and liquid by a factor which optimises at 0.155.
The eventual P-state fraction is 12.8\% in liquid.

A warning is that amplitudes for $^1S_0$, $^3P_1$ and $^3P_2 \to
f_2(1270)$ or $f_2(1565)$ are poorly separated without using information
from interferences between the three $\pi ^0\pi ^0$ contributions,
which will be labelled 12, 23 and 13. The full angular dependence
of amplitudes involves d-matrices which take account of rotations
of axes between 12, 23 and 13 combinations:
\begin {eqnarray}
^3P_1, J_z=+1 &:& 3\cos \alpha \sin \theta \cos \theta -
                  \sin \alpha (3\cos ^2 \theta -1) \\
^3P_1, J_z=0 &:& \sqrt {2}[3\sin \alpha \sin \theta \cos \theta +
                  \cos \alpha (3\cos ^2 \theta -1)] \\
^3P_2, J_z=2 &:& \cos \alpha \sin ^2\theta  -
                  \sin \alpha \sin \theta \cos \theta \\
^3P_2, J_z=1 &:& \sin \alpha \sin ^2\theta  +
                  \cos \alpha \sin \theta \cos \theta .
\end {eqnarray}
For the 12 combination,  $\theta$ is the angle in the $\pi ^0\pi ^0$
rest frame between $\pi _1$ and the recoil $\pi _3$; $\alpha
_{1-3}$ are lab angles of 12, 23 and 13 combinations in their plane.

It turns out that $^3P_2$ annihilation dominates strongly over
$^3P_1$ for both $f_2(1270)$ and $f_2(1565)$.
Visible evidence for $^3P_2 \to f_2(1565)$ is the enhancement
in Fig. 2 near the centre of each band just above 1500 MeV;
this is how the Asterix collaboration discovered $f_2(1565)$
in gas data \cite {Asterix}.
If $^3P_1$ annihilation to $f_2(1270)$ were large, there would be
strong constructive interference between any two crossing bands
near $\cos \theta = 0.6$; there is no sign of
any such enhancement in the data.

\subsection {The $\sigma$ amplitude in $3\pi ^0$ data}
The $\sigma $ is very broad, extending over the full $\pi \pi$ mass
range from 0.27 to 1.74 GeV.
Over such a large range, some $s$-dependence is to be expected
in the numerator $N(s)$ of the production amplitude.
The $s$-dependence required by the data is quite different between
elastic scattering and the production process.
An important feature of the data is an area of low intensity
at the centre of the Dalitz plot of Fig. 2.
Fitting this feature is delicate and demands $s$-dependence in $N(s)$.

Extensive trials have been made using a numerator for the $\sigma$
amplitude of the form
\begin {equation}
N(s) = A[1 + Bs + C/(s + s_0)],
\end {equation}
with $s_0 > 0$.
What emerges is that (a) either complex $B$ or complex $C$ is
definitely required and can fit the data well, (b) including both $B$
and $C$ over-parametrises the amplitude, i.e. strong correlations
develop between $B$, $C$ and $s_0$.
It is better to tolerate a small increase in $\chi^2$, so as to keep
essential features clear with minimal correlations between fitting
parameters.
Final fits were made with the form $A(1 + Bs)$.
Tests were also made including amplitudes arising from the opening
of the $KK$ threshold, i.e. $\propto T_{\pi K}$.
These turned out to be negligible.
A fit was also tried using a dependence on spectator momentum $k$ given
by the diffraction pattern of a black disk, but this gave a poor fit.

\subsection {Prelimiminary remarks on the goodness of fit}
The data are available to me only in the form of binned data,
rather than individual events.
Initial fits revealed that some edge bins have abnormally high
$\chi^2$.
All lie immediately at the edge of the Dalitz plot.
This is actually visible on Fig. 2.
The same effect is observed for $\eta \eta \pi ^0$ data.
The obvious explanation of these bad bins is that the acceptance
near the edge of the Dalitz plot may be incorrectly assessed.
These bad bins have been removed without any significant effect on
fitted amplitudes.

The resulting $\chi^2$ is 2.83 per bin for $3\pi ^0$ data in liquid,
2.85 in gas, 2.64 for $\eta \eta \pi ^0$ in liquid and 2.70
in gas.
A close inspection of discrepancies over the Dalitz plots reveals
only an apparently random scatter, with no clear systematic
effects.
The mean $\chi^2 $ is 2.85 per bin.
To allow for this, all values of $\chi^2$ quoted in the paper are
scaled down by a factor 2.85 so that the average $\chi^2$ becomes
1 per bin.
This is necessary for a correct assessment of the
significance level of observations.

\begin {figure}  [htb]
\begin {center}
\vskip -28mm
\epsfig{file=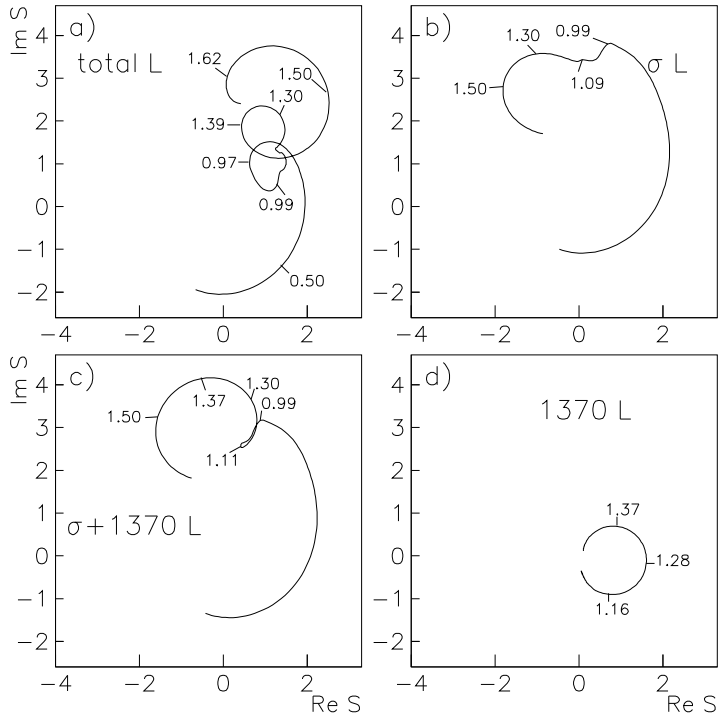,width=11cm}\
\vskip -6mm
\caption{Argand diagrams for the $\pi \pi$ S-wave in liquid hydrogen;
masses are marked in GeV}
\end {center}
\end {figure}

\subsection {Fits including $f_0(1370)$}
The Argand diagram for one $\pi ^0 \pi ^0$
$^1S_0$ combination is shown in Fig. 4(a); individual $\sigma$ and
$f_0(1370)$ components and their coherent sum are shown in other
panels. At low mass, there is a conspicuous loop which is well fitted
with the $\sigma$ pole. This feature was correctly diagnosed by Ishida
et al. \cite {Ishida}. They fitted only the $\pi \pi$ mass projection,
so the present fit is much more accurate.

\begin {figure}  [htb]
\begin {center}
\vskip -14mm
\epsfig{file=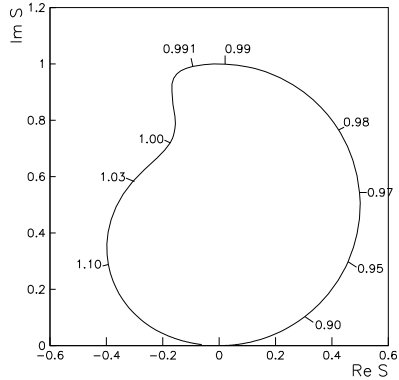,width=6cm}\
\vskip -5mm
\caption{The Argand plot for $f_0(980)$ alone; masses are marked in
GeV}
\end {center}
\end {figure}

At higher mass, there is a loop due to $f_0(980)$.
Here a detail needs explanation.
Fig. 5 shows the Argand diagram for $f_0(980)$ alone in $\pi \pi \to
\pi \pi$.
A mean kaon mass of 495.7 MeV is assumed.
The $KK$ inelasticity sets in very rapidly at threshold, and the
peak inelasticity is at 1.010 GeV.
Thereafter, the inelasticity parameter $\eta$ rises again slowly,
The result is a definite 'dent' in the Argand diagam at 1.01 GeV.

Returning to Fig. 4(a), there is a further loop at $\sim 1300 $ MeV,
followed by a large loop due to $f_0(1500)$.
Fig. 4(a) resembles closely fits made in 1996 \cite {FurtherA}.
The loop at 1300 MeV is the feature which is crucial to the existence
of $f_0(1370)$.
The vital questions are:
\begin {itemize}
\item {(i) is this loop really needed?}
\item {(b) could it be fitted with $\sigma$ and $f_0(1500)$ alone,
without the need for $f_0(1370)$?}
\end {itemize}

\begin {figure}  [htb]
\begin {center}
\vskip -32mm
\epsfig{file=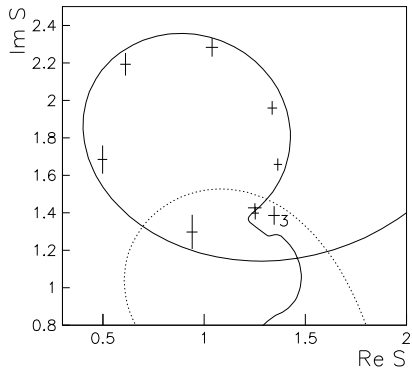,width=12cm}\
\vskip -6mm
\caption{The Argand diagram for the $\pi \pi$ S-wave (full curve)
compared with free fits to magnitude and phase in nine 40 MeV
wide bins of mass from 1100 to 1460 MeV (crosses)}
\end {center}
\end {figure}

\subsection {A suggestion of Ochs}
Ochs has questioned whether the loop at 1300 MeV is an artificial
feature of the way amplitudes are parametrised in terms of resonances
(or backgrounds).
My comment is that the parametrisation must be analytic;
resonance forms used here satisfy this condition.
Nonetheless, he has suggested fitting the $\pi \pi$ S-wave amplitude
freely in magnitude and phase in bins of $\pi \pi$ mass, to see
how definitively the data require the loop at 1300 MeV.

It is not possible to do this over the entire Dalitz plot, because
of strong interferences between one low mass $\pi \pi$ contribution
and two at higher mass.
It is necessary to rely on the strong $\sigma$ loop at low mass and
also the existence and parameters of the $f_0(1500)$, which is well
known today from other data; it is also important to constrain the
$f_0(980)$ within bounds set by BES II data.

It is, however, straightforward to fit the S-wave amplitude freely
in magnitude and phase in bins from 1100 to 1460 MeV, i.e. over the
1300 MeV mass range.
In this test, parameters of $f_0(1500)$ and $f_0(980)$ are allowed
to re-optimise within the narrow ranges allowed by other data.
For the $\sigma$, coefficients $A$ and $b$ of $N(s)$ are set
free, but hardly move because they are determined by elastic data.
Results are shown in Fig. 6 by crosses (indicating errors).
Deviations from the fit of Fig. 4(a), shown by the full curve,
are barely above statistics, except for the third point (labelled 3)
which moves by $\sim 2.5$ standard deviations.
The agreement between the binned fit and Fig. 4(a) rules out the
possibility that the 1300 MeV loop is an artefact of the
parametrisation.
As one example, it is not possible to replace the 1300 MeV loop
by a narrow cusp arising from interference effects; such a cusp
will be illustrated below for the $\pi \pi$ D-wave.

\subsection {Properties of the fitted $f_0(1370)$}
The fitted dispersive contribution $m(s)$ for $f_0(1370)$ is shown
in Fig. 7.
Near 1300 MeV, it varies roughly linearly with $s$, like the term
$(M^2 - s)$ in the Breit-Wigner denominator.
However, $m(s)$ is a factor $\sim 1.6$ larger than $(M^2 - s)$,
This in an unusual feature, showing that care is needed in treating
the $4\pi$ threshold correctly. It also leads to some effects which
had not been forseen.

\begin {figure}  [htb]
\begin {center}
\vskip -15mm
\epsfig{file=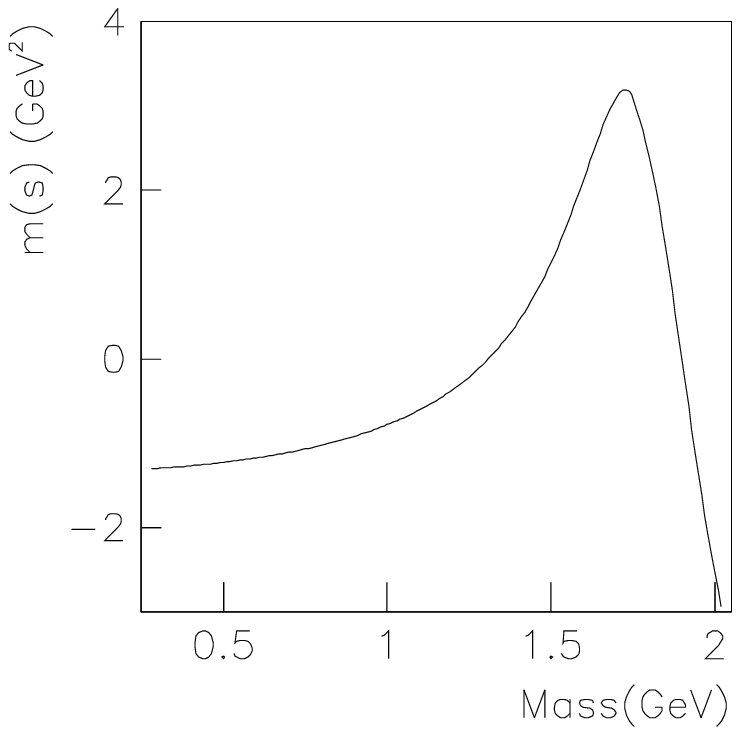,width=6cm}\
\vskip -8mm
\caption{The dispersive term $m(s)$ for $f_0(1370)$}
\end {center}
\end {figure}

The term $m(s)$ is directly linked to $M\Gamma _{4\pi }(s)$ by the
dispersion relation eqn. (2) of Section 1.
This relation is reproducing effects of the loop diagram of Fig. 8.
Results are analogous to vacuum polarisation, and lead to
renormalisation effects in the Breit-Wigner denominator if the
magnitude of $\Gamma _{4\pi}$ is scaled up or down.

\begin {figure}  [htb]
\begin {center}
\vskip -13mm
\epsfig{file=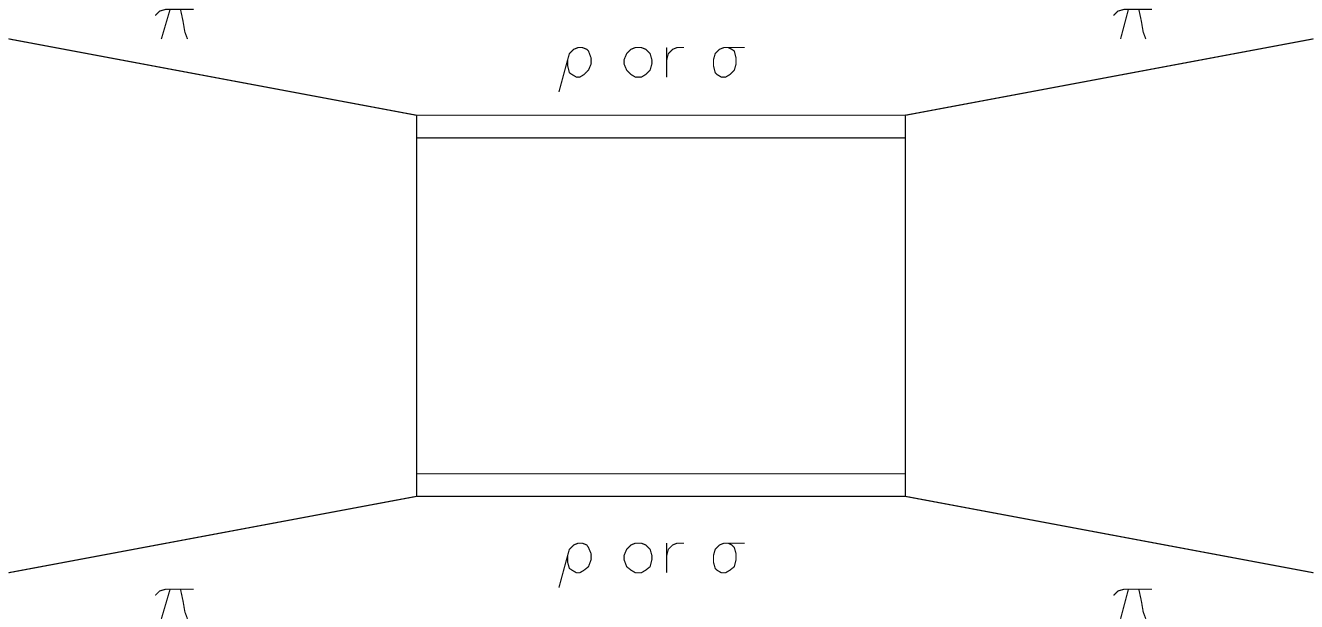,width=7cm}\
\vskip -38mm
\caption{The dispersive term $m(s)$ for $f_0(1370)$}
\end {center}
\end {figure}

A consequence in the combined fit to $\bar pp \to 3\pi ^0$ and
elastic data is that there is a tight constraint on the ratio
$\Gamma _{2\pi}/\Gamma _{4\pi }$.
Conversely, there is considerable flexibility in the absolute
value of $\Gamma _{2\pi }$ which can be fitted.
Table 1 shows pairs of values fitted to data;
$\Gamma _{2\pi }$ and $\Gamma _{4\pi }$ are almost linearly
related until
$\Gamma _{2\pi }$ approaches zero, when the fit
deteriorates rapidly.
The final fit uses the lowest value of $\Gamma _{2\pi }$
giving a satisfactory fit, namely 325 MeV.
A surprise is that $f_0(1370)$ is fairly elastic on resonance,
though the inelasticity increases rapidly thereafter.
Over the range of values shown in Table 1, there is almost no
visible change in the line-shape of $f_0(1370)$.
This is a renormalisation effect. It is illustrated in Fig. 9(b)
for two widely different values of $\Gamma _{2\pi }$.

\begin{table}[htb]
\begin {center}
\begin{tabular}{ccccc}
\hline
$\Gamma _{2\pi}$ & M &  $\Gamma _{4\pi }$ & $\Gamma _{4\pi }/
\Gamma _{2\pi }$ & $\chi ^2$ \\\hline
0.80 & 1.3113 & 0.1958 & 0.245 & 3519 \\
0.65 & 1.3090 & 0.1472 & 0.226 & 3507 \\
0.50 & 1.3093 & 0.1047 & 0.209 & 3505 \\
0.40 & 1.3093 & 0.0766 & 0.192 & 3502 \\
0.30 & 1.3096 & 0.0464 & 0.155 & 3500 \\
0.25 & 1.3089 & 0.0318 & 0.127 & 3504 \\
0.20 & 1.3114 & 0.0181 & 0.091 & 3512 \\\hline
\end{tabular}
\caption{Parameter variations with $\Gamma _{2\pi}$ of $f_0(1370)$;
units are GeV.}
\end{center}
\end{table}

\begin {figure}  [htb]
\begin {center}
\vskip -19mm
\epsfig{file=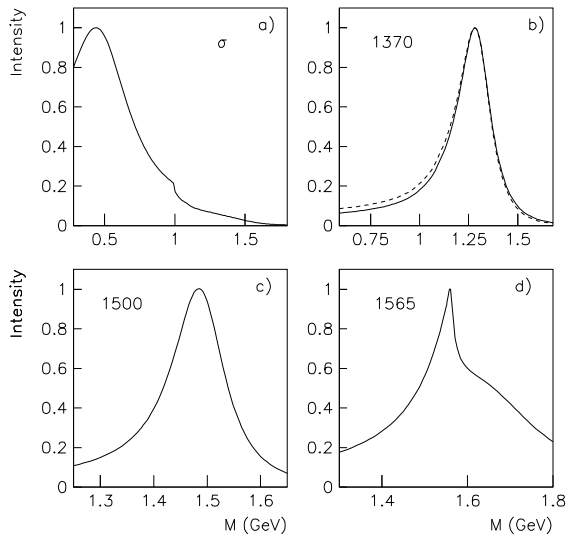,width=8.5cm}\
\vskip -4.5mm
\caption{Line-shapes of $\sigma$, $f_0(1370)$, $f_0(1500)$ and
$f_2(1565)$ in $\pi \pi \to \pi \pi$, normalised to 1 at their peaks;
the dashed curve for $f_0(1370)$ shows the effect of changing
$\Gamma_{2\pi}$ from 325 to 800 MeV}
\end {center}
\end {figure}

\subsection {Lineshapes}
Fig. 9 shows the line-shapes of $\sigma$, $f_0(1370)$, $f_0(1500)$
and $f_2(1565)$ in $\pi \pi$ elastic scattering.
The $f_0(1370)$ is almost degenerate with $f_2(1270)$, which
explains why it has been hard to observe experimentally.
The peak of the intensity is at 1282 MeV in
elastic scattering, though the phase shift goes through $90^\circ$
at $M = 1309$ MeV.
A similar pole mass is quoted by Anisovich and Sarantsev \cite
{AandS02}: $1306 - i147$ MeV for their solution 1 and marginally
different for two alternative solutions.
Although their parametrisation of amplitudes in terms of $K$-matrices
is quite different to the approach adopted here, it is clear that
their eventual fit is very similar to mine, and there is
no serious disagreement between the two types of formalism.

The $f_0(1370)$ and $f_0(1500)$ are asymmetric, because the rapidly
rising $\Gamma _{4\pi}$ in the Breit-Wigner denominator cuts off
the intensity at high mass.
The $f_2(1565)$ is likewise cut off strongly at high mass by the
opening of the $\omega \omega$ threshold.
It peaks exactly at the $\omega \omega$ threshold and has a
half-width of 131 MeV below resonance.

\begin {figure}  [htb]
\begin {center}
\vskip -28mm
\epsfig{file=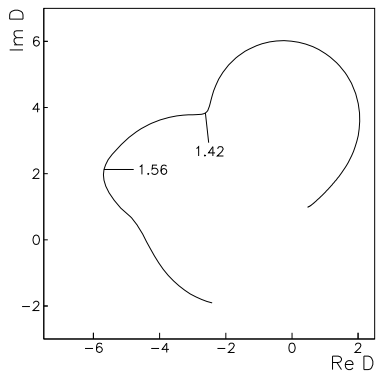,width=5.5cm}\
\vskip -3mm
\caption{The Argand plot for the $\pi \pi $ D-wave in $3\pi ^0$ data
in liquid hydrogen; masses are shown in GeV}
\end {center}
\end {figure}

\subsection{The $\pi \pi $ D-wave}
Fig. 10 shows the Argand diagram for $^1S_0$ annihilation to the
$\pi \pi $ D-wave.
An interesting feature, already noted in Ref. \cite {FurtherA}, is
the appearance of a cusp at 1420 MeV, almost midway between
$f_2(1270)$ and $f_2(1565)$.
No resonance can be accomodated at 1420 MeV by fitted magnitudes and
phases of the amplitude.
However, the PDG lists a state $f_2(1430)$.
It seems likely that this is an artefact due to similar cusps
in other reactions; it is an acute observation on the part of
the experimental groups finding the effect.

\subsection {P-state annihilation}
Fig. 11 shows Argand diagrams for $^3P_1$ annihilation
to one $\pi \pi$ S-wave combination.
The result comes almost purely from data in gas.
There is some similarity to results in liquid, but definite
differences; for example, the $f_0(980)$ contribution is almost
negligible.
There is again a loop near 1300 MeV in Fig. 11(b).
However, interference between $f_0(1370) $ and $f_0(1500)$ plays
a decisive role in obtaining a good fit.

\begin {figure}  [htb]
\begin {center}
\vskip -16mm
\epsfig{file=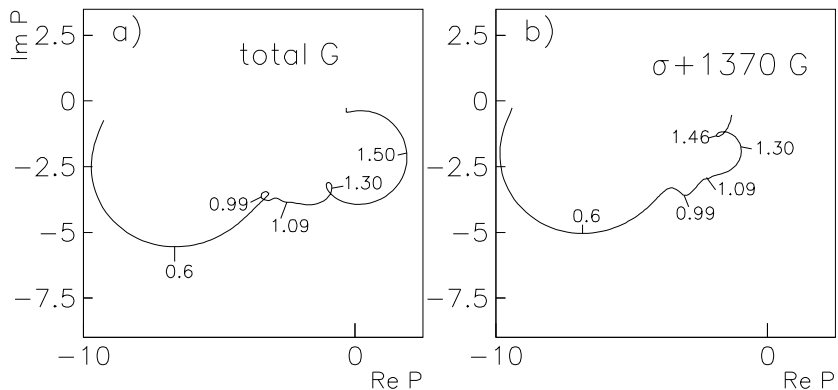,width=12.5cm}\
\vskip -6mm
\caption{Argand diagrams for the $\pi \pi$ S-wave in $^3P_1$;
annihilation;
masses are shown in GeV}
\end {center}
\end {figure}

\begin {figure}  [htb]
\begin {center}
\vskip -20mm
\epsfig{file=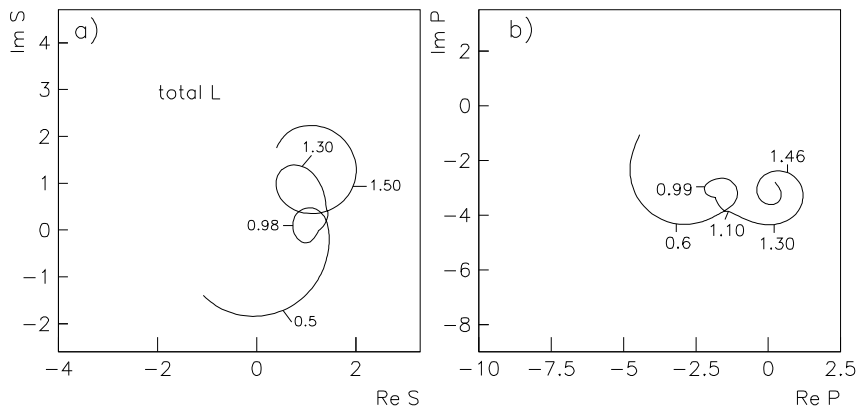,width=13cm}\
\vskip -8mm
\caption{Argand plots for the $\pi \pi$ S-wave in $3\pi ^0$ data in
liquid (L) and gas (G) with $f_0(1370)$ omitted; masses are shown in
GeV}
\end {center}
\end {figure}

\subsection {Fits without $f_0(1370)$}
The entire fit has been re-optimised without $f_0(1370)$, refitting
all parameters.
The result is that (renormalised) $\chi ^2$ increases for $3\pi ^0$
data in liquid by 1040, and by 1088 for data in gas.
So the significance level of $f_0(1370)$ is slightly over
32 standard deviations in liquid and close to 33 in gas.
These are essentially independent, since $^1S_0$ annihilation
to $f_0(1370)$ is determined almost purely by liquid data and
$^3P_1$ annihilation by gas data.

Attempts were made to improve the fit using different parametrisations
for the $\sigma$, e.g. $N(s) \propto A[1 + C/(s + s_0)]$; however,
that choice actually gave a slightly worse fit and other types of fit
gave no significant improvement.

Fig. 12 shows the resulting Argand diagrams for the $\pi \pi$ S-wave
in (a) liquid, (b) gas.
The fit tries to remedy the situation by using the $\sigma$ amplitude
as a replacement for $f_0(1370)$; the main latitude lies in increasing
the inelasticity of $\sigma$ to $4\pi$.
This fails because the amplitude cannot move round the required
loop quickly enough, resulting in a very poor $\chi^2$;
one can see on Fig. 12 that the size of the loop at 1300 MeV has
increased compared with that of Fig. 4.
For the $\sigma $ amplitude, the inelasticity to $4\pi$ rises slowly
over the entire mass range 1200 to 2100 MeV.
This produces a slow loop
on the Argand diagram, see Fig. 4(b) for liquid(L).
It is not possible to fit the narrow $f_0(1370)$, whose full-width at
half-maximum is $207$ MeV, with this slow loop.

Tornqvist \cite {Tornqvist} has remarked that the dispersive
contribution $m(s)$ to the $\sigma$ amplitude could induce a
second $\sigma$ pole near the $4\pi$ threshold.
All fits have been examined for such a pole, but there is no trace
of it in any fit.

The conclusion is that a narrow $f_0(1370)$ is highly significant.
However, one should not rely purely on $\chi^2$.
What adds considerable confidence is that fitted values of mass
and width are in excellent agreement between two almost
independent determinations in liquid data and gas.
The value of $\Gamma _{2\pi }$ is held fixed at 325 MeV in both
cases. Then the fitted value of $\Gamma _{4\pi}$ on resonance
changes by only 4 MeV.
Parameters are shown in Table 2.
One must add a systematic error common to both determinations;
the systematic errors shown in Table 2 cover the entire range of
all observed fits to the six sets of data with any parametrisation
of $N(s)$ for the $\sigma$ amplitude.

\begin{table}[htb]
\begin {center}
\begin{tabular}{ccccc}
\hline
 & (a) & (b) & (c) & (d) \\\hline
 $\Gamma _{2\pi}(M^2) $ & 325 & 325 & 325 & 127 \\
M & 1308 & 1312 & $1309 \pm 1(stat) \pm 15(syst)$ &
$1503 \pm 1(stat) \pm 6(syst)$ \\
$\Gamma _{4\pi }(M^2)$ & 53 & 56 & $54 \pm 2(stat) \pm 5(syst)$ &
$138 \pm 4(stat) \pm 5(syst)$ \\
peak & & & 1282 & 1485 \\
half-height & & & 1165 and 1372 & 1418 and 1540 \\
FWHM & & & 207 & 122 \\\hline
\end{tabular}
\caption{Parameters in MeV for $f_0(1370)$ in
(a) liquid, (b) gas, (c) combined, and $f_0(1500)$ for the combined
fit.}
\end{center}
\end{table}

\begin {figure}  [htb]
\begin {center}
\vskip -32mm
\epsfig{file=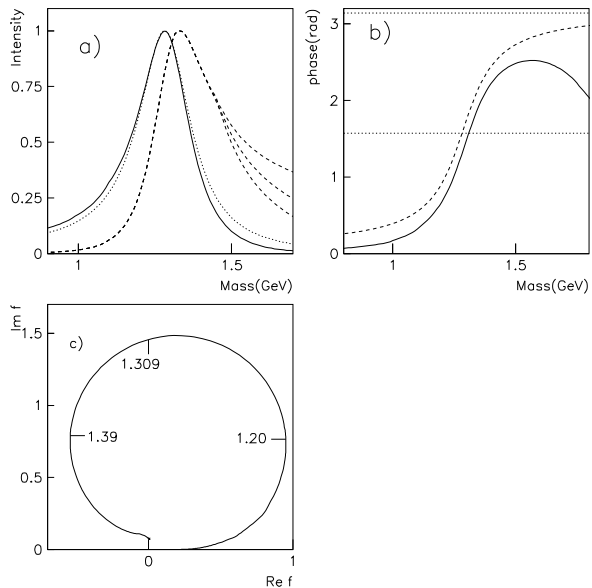,width=9cm}\
\vskip -8mm
\caption{(a) line-shapes of $f_0(1370)$ for $2\pi$ (full curve), a
Breit-Wigner amplitude with constant width (dotted), and for $4\pi$
(dashed), (b) the phase angle measured from the bottom of the
Argand plot (full curve) and for a Breit-Wigner amplitude of constant
width (dashed); horizontal lines mark phase shifts of $\pi/2$ and
$\pi$,  (c) Argand plot; masses are shown in GeV.}
\end {center}
\end {figure}

For those who wish to reproduce $f_0(1370)$ and $f_0(1500)$ with
less elaborate formulae than used here, the advice is to aim to
reproduce the peak and half-heights of Table 2.

A further test has been made removing the phase variation of
$f_0(1370)$.
There is an increase in $\chi^2 $ of 165, i.e. nearly 13 standard
deviations. However, the magnitude of the fitted $f_0(1370)$
increases by a large amount and the $\sigma$ component decreases.
So the non-resonant $f_0(1370)$ is then obviously simulating a
large part of the $\sigma$ amplitude.
This confuses the interpretation of this test.

\subsection {$f_0(1370)$ line-shape in $4\pi$}
Fig. 13(a) shows as the full curve the line-shape of $f_0(1370)$
in $\pi \pi \to \pi \pi$; the dotted curve shows the line-shape of
a simple Breit-Wigner amplitude of constant width with the same
mass and width on resonance.
The dashed curves show estimates of what is predicted in
$\pi \pi \to 4\pi$.
There is some uncertainty concerning form factors
at high mass.
Fortunately these form factors play little role in fitting
present data, because the amplitude in the $2\pi$ channel is
already quite small at 1.45 GeV, where the form factor begins
to have an effect.
The three curves illustrate results using form factors
$\exp -\alpha (s - 1.45^2)$ (with $s$ in GeV$^2$); the top curve is
for $\alpha = 0.5$, the central one for $\alpha = 1.0$ and the
bottom one for $\alpha = 1.5$.
These span a reasonable range of possibilities.

The main conclusion from these curves is that the peak of
the resonance moves quite strongly between $2\pi$ and $4\pi$
channels because of the difference between $\rho _{2\pi }$
and $\rho _{4\pi }$.
The $2\pi$ peak is at 1282 MeV and that in $4\pi$ is at 1331
MeV, i.e. a mass difference of $\sim 50$ MeV.
However, if one works from the half-heights of the peaks, the
difference is larger.
For the $2\pi $ channel, half-height is at 1165 and 1372 MeV,
i.e. a mean of 1269 MeV and a full-width at half maximum (FWHM) of
207 MeV.
For the $4\pi$ channel, the corresponding half heights are
1241 and 1514 MeV, i.e. a mean of 1377 MeV and FWHM = 273 MeV.
Further data on $\pi \pi \to 4\pi$ would be very valuable.
In particular, data separating spins 0 and 2 would help greatly in
clarifying the parametrisation of $f_2(1565)$. However, the analysis of
these data must take into account contributions from $\sigma \to 4\pi$;
that has not been done up to the present.

Fig. 13(c) shows the Argand loop for $f_0(1370)$ in elastic
scattering.
The loop is cut off at the left by the effect of
$4\pi$ inelasticity.
It is remarkable that the loop is very close in shape to the
circle given by a Breit-Wigner resonance of constant width.
This provides some support for the constant width approximation
which is frequently used.
The resonance is behaving to first approximation as a simple
pole with appropriate widths to $2\pi$ and $4\pi$.
However, Fig. 13(b) compares the phases for the $s$-dependent
form of the amplitude (full curve) and constant width (dashed);
these phases are measured from the origin of the Argand diagram.
There is a sizable difference in phases, but only above
the upper half-width of the resonance.

\subsection {Parametrisation of $m(s)$}
For convenience, an algebraic parametrisation of
$m(s)$ is given here, to allow reconstruction of
the amplitudes for $f_0(1370)$, $f_0(1500)$ and $\sigma$.
It is not possible to find a simple accurate formula
dealing with all three mass ranges.
Instead, formulae will be given which are sufficiently
accurate for extrapolations from the physical region to
the poles.
This implies weighting the fit to $m(s)$ in the vicinity
of these poles.

The form of parametrisation is guided by the facts that
(a) $\Gamma _{4\pi}$ is parametrised as a Fermi function,
(b) the dispersive term is given approximately by the
gradient of this function.
Then $m(s)$ is expressed as a sum of terms
\begin {equation}
m(s) = \sum _i \left( \frac {a_i}{(s - s_i)^2 + w_i^2} -
\frac {a_i}{(M^2_i - s_i)^2 + w_i^2} \right).
\end {equation}
Table 3 gives numerical values of parameters.
Note that the
$\sigma$ parametrisation applies only to the vicinity of the pole;
if the full form of $m(s)$ is needed for the $\sigma$ over
the entire mass range, values may be read from Fig. 7 or the
author will supply numerical tables.

\begin{table}[htb]
\begin {center}
\begin{tabular}{cccc}
\hline
 & $f_0(1370)$ & $f_0(1500)$ & $\sigma $ \\\hline
$M$   & 1.3150 & 1.5028 & 0.9128 \\
$a_1$ & 3.5320  & 1.4005 & 17.051 \\
$s_1$ & 2.9876  & 2.9658 &  3.0533 \\
$w_1$ & 0.8804  & 0.8129 &  1.0448 \\
$a_2$ & -0.0427 & -0.0135 & -0.0536 \\
$s_2$ & -0.4619 & -0.2141 & -0.0975 \\
$w_2$ & -0.0036  & 0.0010 &  0.2801 \\\hline
\end{tabular}
\caption{Parameters fitting $m(s)$ in units of GeV.}
\end {center}
\end{table}

A full account of pole positions on the many possible sheets
is not helpful.
The sheets may be labelled by the sign multiplying $i$.
Table 4 then lists a representative set for $f_0(1370)$ and $f_0(1500)$.
One sees immediately that the sign attached to $i$ for $KK$ and $\eta
\eta$ has little effect on the pole position, because these
inelasticities are small.
What matters are the signs of $i$ for $\pi \pi$ and $4\pi$ sheets.
The imaginary part of the pole position changes
substantially when the sign for the $4\pi$ sheet changes.
This is a familiar effect of a strong inelastic channel.
The experimental line width is close to the average of the results
for the two $4\pi$ sheets.

\begin{table}[htb]
\begin {center}
\begin{tabular}{cccccl}
\hline
State & $\pi \pi$ & $4\pi$ & $KK$ & $\eta \eta$ & Pole (MeV) \\\hline
$f_0(1370)$ & + & + & + & + & $1299 -i187$ \\
            & + & - & + & + & $1309 -i43$ \\
            & + & + & - & + & $1293 -i180$ \\
            & + & + & + & - & $1292 -i177$ \\
$f_0(1500)$ & + & + & + & + & $1492 - i104$ \\
            & + & - & + & + & $1497 - i53$ \\
            & + & + & - & + & $1492 - i103$ \\
            & + & + & + & - & $1492 - i103$ \\\hline
\end{tabular}
\caption{Pole position on various sheets.}
\end {center}
\end{table}

\begin {figure}  [htb]
\begin {center}
\vskip -28mm
\epsfig{file=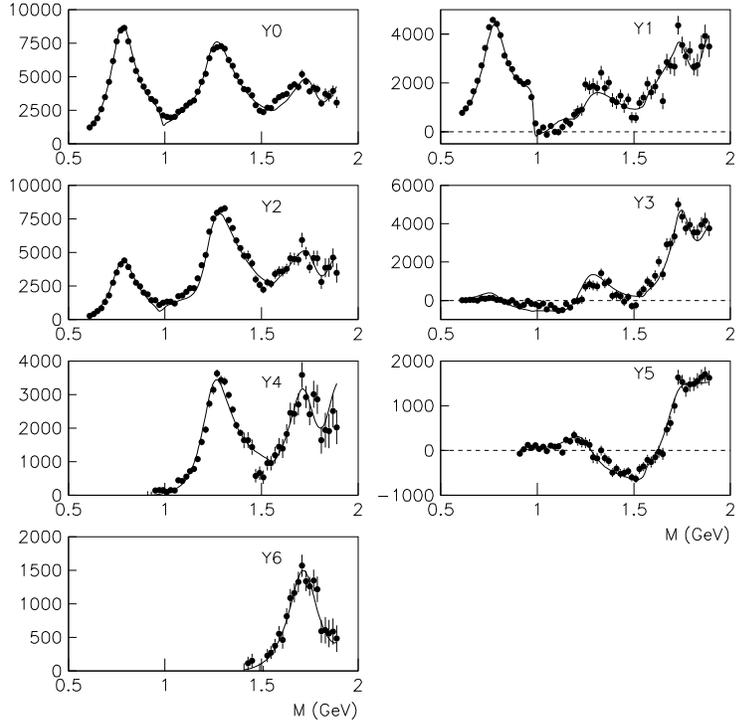,width=11cm}\
\vskip -8mm
\caption{Fit to Cern-Munich moments with $M=0$}
\end {center}
\end {figure}

\section {Fits to elastic scattering}
Four sources of information on elastic scattering are fitted
simultaneously with production data.
The first are Cern-Munich data from $\pi p \to (\pi \pi) n$.
Secondly, $K_{e4}$ data of Pislak et al. are included and constrain
the $\pi \pi$ S-wave phase shifts up to 382 MeV \cite {Pislak}.
Thirdly, Caprini et al have made a prediction of $\pi \pi$ phases
using the Roy equations \cite {Caprini}.
Their prediction up to 925 MeV is included with errors which are
adjusted to give $\chi^2 = 1$ per point.
These predictions are particularly important in constraining the
scattering length and effective range.
Amplitudes for $\pi\pi$ isospin 2 amplitudes are also included,
and parametrisations are given below.
Fourthly, BES II data on $J/\Psi \to \omega \pi^+\pi ^-$
provide an accurate parametrisation of the $\sigma$ pole, as discussed
in Ref. \cite {sigpole}.
The prediction of the lower side of the $\sigma$ pole from the Roy
equations is precise, but the BES data determine the upper side more
accurately because of effects arising from the sub-threshold $KK$ and
$\eta \eta$ contributions.

In fitting Cern-Munich data, $J^P = 0^+$ contributions are included
from $\sigma$ and $f_0$'s at 980, 1370, 1500, 1790, and 2020
MeV; the last of these is above the mass range of data, which finish
at 1.89 MeV, but it needs to be included because of its large
width.
The $f_0(1710)$ is dominantly $s\bar s$ and is not expected to
contribute strongly; any possible contribution is absorbed into the
parameters of $f_0(1790)$.

Contributions are allowed for $J^P = 1^-$ from $\rho$'s at 770,
1450, 1700, 1900 and 2000 MeV, although the $\rho (2000)$
is included only for completeness.
For spin 2, $f_2(1270)$ plays a dominant role, but $f_2(1565)$ is
definitely needed, as is some contribution from either or both of
$f_2(1920)$ and $f_2(1950)$; the latter two however, cannot be
separated cleanly. The $I = 2$ D-wave is also included, with
formulae discussed in the next sub-section.

For spin 3, $\rho (1690)$ plays a strong role, but there is definite
evidence for some additional contribution from $\rho _3(1990)$.
Finally, some definite but small contribution from $f_4(2040)$ is
needed.

Before going into details, final fits to Cern-Munich moments
are shown in Figs. 14 and 15.
Panels are labelled by L,M of spherical harmonics fitted to data.
The fit is quite adequate, but the eventual $\chi ^2$ is 3.13 per
point.
There is, for example, a definite systematic discrepancy with the
$Y(51)$ moment, Fig. 15.
This shows structure around 1270 MeV which
cannot reasonably be attributed to interference with the low mass tail
of $\rho (1690)$: the effect is too large. Near 1550 MeV, there are
discrepancies with Y2 and Y4 moments, probably because of the effect
of the $\rho \rho$ threshold on $f_2(1565)$; this is not explicitly
included.

\begin {figure}  [htb]
\begin {center}
\vskip -16mm
\epsfig{file=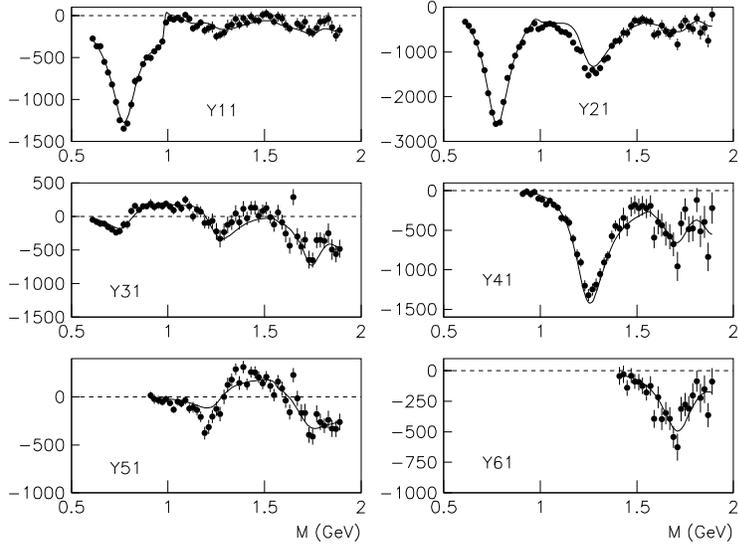,width=11cm}\
\vskip -28mm
\caption{Fit to Cern-Munich moments with $M=1$}
\end {center}
\end {figure}

\begin {figure}  [htb]
\begin {center}
\vskip -28mm
\epsfig{file=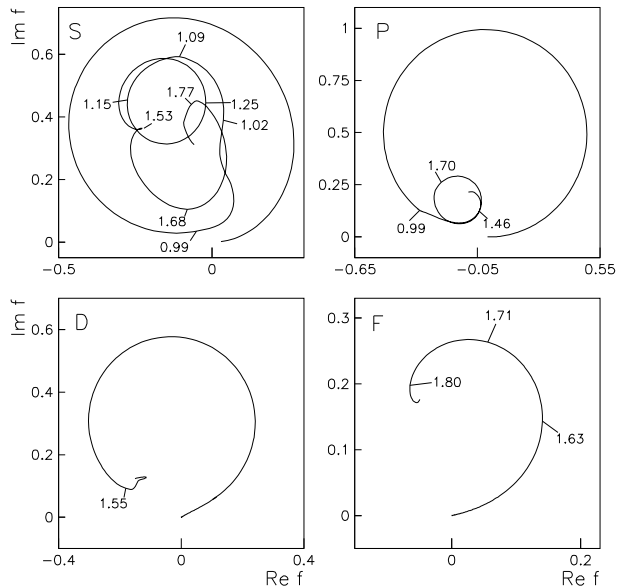,width=9cm}\
\vskip -6mm
\caption{Argand diagrams for $\pi \pi$ partial waves in elastic
scattering}
\end {center}
\end {figure}

The fit to $f_0(980)$ is constrained within the linear
sum of statistical and systematic errors quoted by BES for its
mass, $g^2(\pi \pi)$ and $g^2(KK)/g^2(\pi \pi)$.
The last of these is a useful constraint, but the fit does
optimise the other two parameters within the BES errors, showing
there is no conflict with Cern-Munich and Crystal Barrel data.
The mass and width of the prominent $\rho(770)$ also need to be
fitted freely, with the results $M = 778$ MeV, $\Gamma = 153$ MeV,
in satisfactory agreement with PDG values.
A small detail is that its coupling to $4\pi$ is included using on
resonance the PDG estimate of the branching ratio; including this
effect marginally improves the fit to the tail of the $\rho$ above
1 GeV.

\subsection {$I=2$ amplitudes}
There are experimental data on $\pi ^+ p \to (\pi ^+ \pi ^+)n$
and hence the $I = 2$ S-wave amplitude.
Wu et al.  have calculated the expected inelasticity \cite
{WuZou}.
These two inputs have been fitted empirically as follows:
\begin {eqnarray}
f(I=2)&=& (\eta \exp (2i\delta) - 1)/2i, \\
\eta &=& 1.0 - 0.5 \rho _{4\pi} (s),  \\
\tan \delta &=& \frac {12.9 \rho_{2\pi }}{11.9 + s/m^2_\pi}
(a + bq^2 + cq^4 + fq^6),
\end {eqnarray}
where $q$ is pion momentum in the $\pi \pi$ rest frame,
and a = -0.0444, b=-0.12508, c = -0.00561, d = 0.00014,
all in units of pion masses.

The I=2 D-wave has been studied carefully by Pelaez \cite {Pelaez}.
His formula for the phase shift $\delta$ is used.
The elasticity $\eta$ is set to 1 up to 1.05 GeV, and thereafter
taken from the work of Wu et al. \cite {WuZou} as
\begin {equation}
\eta = 1 - 0.2(1.0 - 1.05^2/s)^3.
\end {equation}
The effect of this amplitude is very small in the present analysis.

\begin {figure}  [htb]
\begin {center}
\vskip -32mm
\epsfig{file=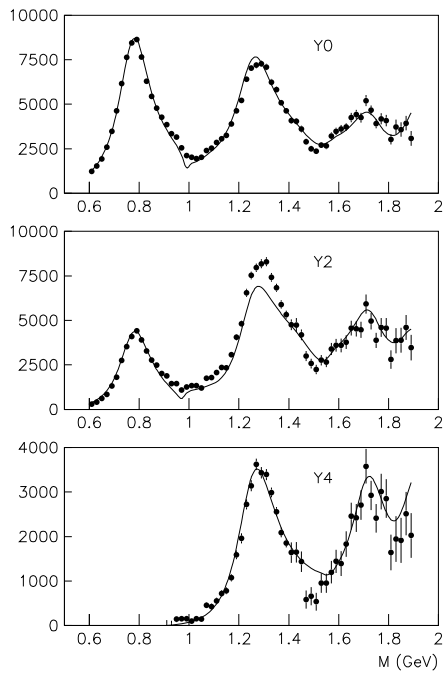,width=7cm}\
\vskip -6mm
\caption{The fit to Cern-Munich moments without mixing between
$\sigma$, $f_0(1370)$ and $f_0(1500)$}
\end {center}
\end {figure}

\begin {figure}  [htb]
\begin {center}
\vskip -16mm
\epsfig{file=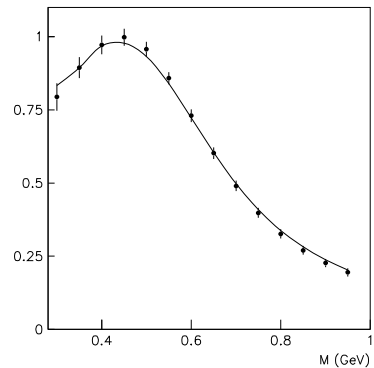,width=6cm}\
\vskip -6mm
\caption{The fit to the $\pi \pi$ mass projection of BES II data for
$J/\Psi \to \omega \pi ^+\pi ^-$}
\end {center}
\end {figure}

\subsection {Comments on fitted resonances}
The $\pi \pi$ S-wave plays an important role in fitting
moments with $L = 0$, 1, 2 and 3.
The value of $\Gamma _{2\pi}$ for $f_0(1500)$ is well determined and
significantly larger than has been generally assumed.
Values of $\Gamma _{2\pi}$ are collected into Table 5.
Values of $\chi^2$ are renormalised downwards by a factor 3.13 in order
to allow for the high $\chi^2$ of the fit.
The value of $\Gamma _{2\pi}$ for $f_0(1790)$ is not well
determined because of overlap with the broad $f_0(2020)$.
Argand diagrams are shown in Fig. 16.

Without $f_0(1370)$, a different type of solution can be obtained,
but without the extra loop of Fig. 16 between 1.15 and 1.25 GeV.
If the mass and width of $f_0(1500)$ are fitted freely, the $\chi^2$
without $f_0(1370)$ is worse by 26.
However, the width of $f_0(1500)$
goes up considerably, obviously because it is simulating missing
$f_0(1370)$; also the mass goes down. If both mass and width of
$f_0(1500) $ are held fixed, $\chi^2 $ is worse by 45 than with
$f_0(1370)$ included.

Either or both of $\rho (1450)$ and $\rho (1700)$ are required
to reproduce $L = 1$ moments, but are poorly separated.
One can remove either with a change in $\chi ^2 < 10$.
The mass and width of $\rho(1450)$ are taken from Babar data
\cite {BABAR}, where a conspicuous $a_1(1260)\pi$ signal is observed.
The $\rho (1900)$ is fixed in width to the value 145 MeV,
the mean of the two sets of Babar data \cite{Babrho}.
It gives a significant improvement, but is
not well separated from $\rho (2000)$.
It may well be a radial excitation of $\rho (1450)$, but could
also be a cusp effect due to the opening of the $\bar pp$
threshold. A fit using the narrow width of Frabetti et al
\cite {Frabetti} is somewhat poorer.
This raises the possibility of a resonance which has been
attracted to the $\bar pp$ threshold; Frabetti et al may be
sensitive to the threshold effect, while Babar data may be
more sensitive to the resonance.

Spin 2 states contribute strongly to moments with L = 2 to 5.
The $f_2(1565) $ is definitely required but cannot be parametrised
accurately because of missing information on $\Gamma _{4\pi}$.

The $\rho _3(1690)$ makes a strong contribution to moments with
$L = 5$ and 6.
Its optimum mass is $1709 \pm 6$ MeV.
Reducing it to the PDG average of $1688.7 \pm 2.1$ does not
affect the fit to $L=6$ moments significantly, but does
make the fit to $L = 5$ moments worse by 20 in $\chi^2$.
The value of $\Gamma _{2\pi}/\Gamma _{tot}$ is $0.224 \pm 0.032$,
in close agreement with the PDG value.

\begin{table}[htb]
\begin {center}
\begin{tabular}{cccc}
\hline
State & $\Gamma _{2\pi}$ (MeV) & $\Delta \chi ^2$  \\\hline
$f_4(2040)$ & $16 \pm 7$ & 5.3 & \\
$\rho _3(1990)$ & $36 \pm 8$ & 12.4 \\
$\rho _3(1690)$ & $44 \pm 3$ & 497 \\
$f _2(1565)$ & $46 \pm 14$ & 33 \\
$f _0(1500)$ & $61 \pm 5$ & 149\\\hline
\end{tabular}
\caption{Parameters of fitted resonances; column 3 shows the
(renormalised) change in $\chi ^2 $ when the resonance is removed
from the fit to Cern-Munich data.}
\end {center}
\end{table}

The fits shown in Figs. 14 and 15 include mixing between $\sigma$,
$f_0(1370)$ and $f_0(1500)$. Fig. 17 shows the fit to $Y0$, $Y2$, and
$Y4$ with the mixing removed. Without this mixing, the fit to $Y2$ is
not satisfactory, because interferences between $f_2(1270)$ and
$f_0(1370)$ are not accurately fitted. Extra phase variations from the
mixing play an important role. Such mixing is to be expected for
strongly overlapping resonances.

Fig. 18 shows the fit to the $\pi \pi$ mass projection of
BES data for $J/\Psi \to \omega \pi ^+\pi ^-$, normalised to 1 at
the highest data point.
The $\sigma$ pole is at $470 \pm 30 - i(260 \pm 30)$ MeV.
These values have changed little from the work of Ref. \cite {sigpole}.
Parameters fitted to the $\sigma$ amplitude are shown in Table 6
in units of GeV.
A parametrisation of $m(s)$ in the vicinity of the $\sigma$
pole is shown above in Table 3.

\begin{table}[htb]
\begin {center}
\begin{tabular}{cccc}
\hline
Parameter & Value\\\hline
$M$ & 0.900\\
$b_0$ & 3.728\\
$b_1$ & 0.092\\
$A$ & 2.882 \\
$g_4$ & 0.012\\\hline
\end{tabular}
\caption{Parameters fitted to the $\sigma$ amplitude in units of GeV}
\end {center}
\end{table}

\begin {figure}  [h]
\begin {center}
\vskip -28mm
\epsfig{file=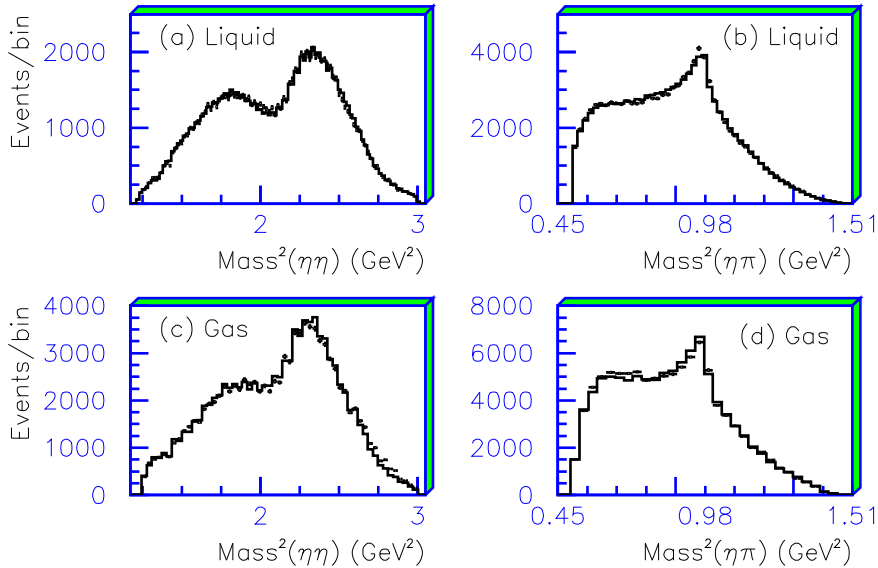,width=12cm}\
\vskip -6mm
\caption{Mass projections from data on $\bar pp \to \eta \eta \pi
^0$ at rest in liquid hydrogen and gas; points show data and the
histogram shows the fit}
\end {center}
\end {figure}

\begin {figure}  [htb]
\begin {center}
\vskip -18mm
\epsfig{file=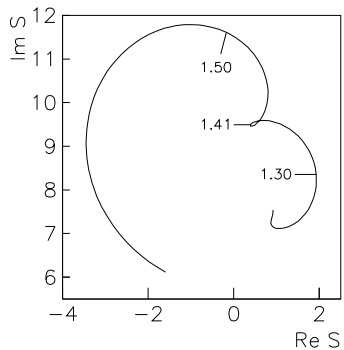,width=10cm}\
\vskip -46mm
\caption{The Argand diagram for the $\eta \eta$ S-wave in
$\eta \eta \pi ^0$ data in liquid; masses are shown in GeV.}
\end {center}
\end {figure}

\section {Fits to $\bar pp \to \eta \eta \pi ^0$ data}
The $\eta \pi$ and $\eta \eta$ mass projections are shown in Fig.
19 for data in liquid hydrogen and gas.
In (b) and (d) there are obvious peaks due to $a_0(980)$.
In (a) and (c) there is a strong peak due to $f_0(1500)$
and a lower peak which is naturally attributed to $f_0(1370)$;
it cannot be due to $f_2(1270)$, whose coupling to $\eta \eta$ is
much too weak.

The ingredients in the fit are $a_0(980)$ and $a_2(1320)$,
$\sigma$, $f_0(1370)$, $f_0(1500)$, $f_2(1270)$ and $f_2(1525)$.
The $f_2(1270) \to \eta \eta$ contributes very little intensity,
because of its small branching ratio, but it is desirable to
include it in order to accomodate possible interference effects.
This is done, fixing the branching ratios between $\eta \eta \pi ^0$
and $3\pi ^0$ data according to the PDG value for
$BR[f_2(1270) \to \eta \eta ]/BR[f_2(1270) \to \pi \pi]$.

An earlier Crystal Barrel paper \cite {AmslerD} has noted that
there is evidence for $f_2(1525) \to \eta \eta$. However, its
branching ratio is sufficiently low that it would not be seen
in $3\pi ^0$ data.
This result is confirmed here, and the $f_2(1525)$ makes a large
improvement to fitting $\eta \eta \pi$ data, $\Delta \chi^2 = 344$
summed over data in liquid and gas.

Additional fits have been made including the exotic $\pi _1(1400)$
in the $\pi \eta$ P-wave; however, it makes no significant
contribution.
The $a_0(1450)$ is above the available mass range in $\eta \pi$, but
its low mass tail could contribute and this has been tried.
However, the fits are no better than can be obtained
by including form factors into the high mass tail of $a_0(980)$ in
$\eta \pi $ and $KK$ channels.
Incidentally, the decay channel
$a_0(980) \to \eta ' \pi$ is included, assuming that the
ratio $g^2[a_0(980) \to \eta ' \pi]/g^2[a_0(980) \to \eta \pi
] = (0.6/0.8)^2$, as predicted by the pseudoscalar mixing angle.
This does improve the fit significantly, but needs further study for
$\bar pp \to \eta \pi ^0 \pi ^0$, where the $a_0(980)$ signal is
more distinct.

The Argand diagram fitted to the $\eta \eta$ S-wave is
shown in Fig. 20. There is a distinct cusp between $f_0(1370)$
and $f_0(1500)$. Masses and widths fitted to both resonances are
entirely consistent with $3\pi ^0$ data.

Without $f_0(1370)$ in the fit, the $\sigma$ amplitude makes some
contribution to replacing it. However, $\chi ^2$ is worse by
317 for liquid data, i.e. $>17$ standard deviations, and 68 for data
in gas, $>8 $ standard deviations.
The significance level in liquid is high and confirms
that the peak at $\sim 1300$ MeV in $\eta \eta$ cannot be fitted
adequately with the $\sigma $ contribution alone.
This result by itself is sufficient to show the existence of
$f_0(1370)$, but the determination of its mass and width are much
poorer than from $3\pi ^0$ data.

\begin {figure}  [h]
\begin {center}
\vskip -28mm
\epsfig{file=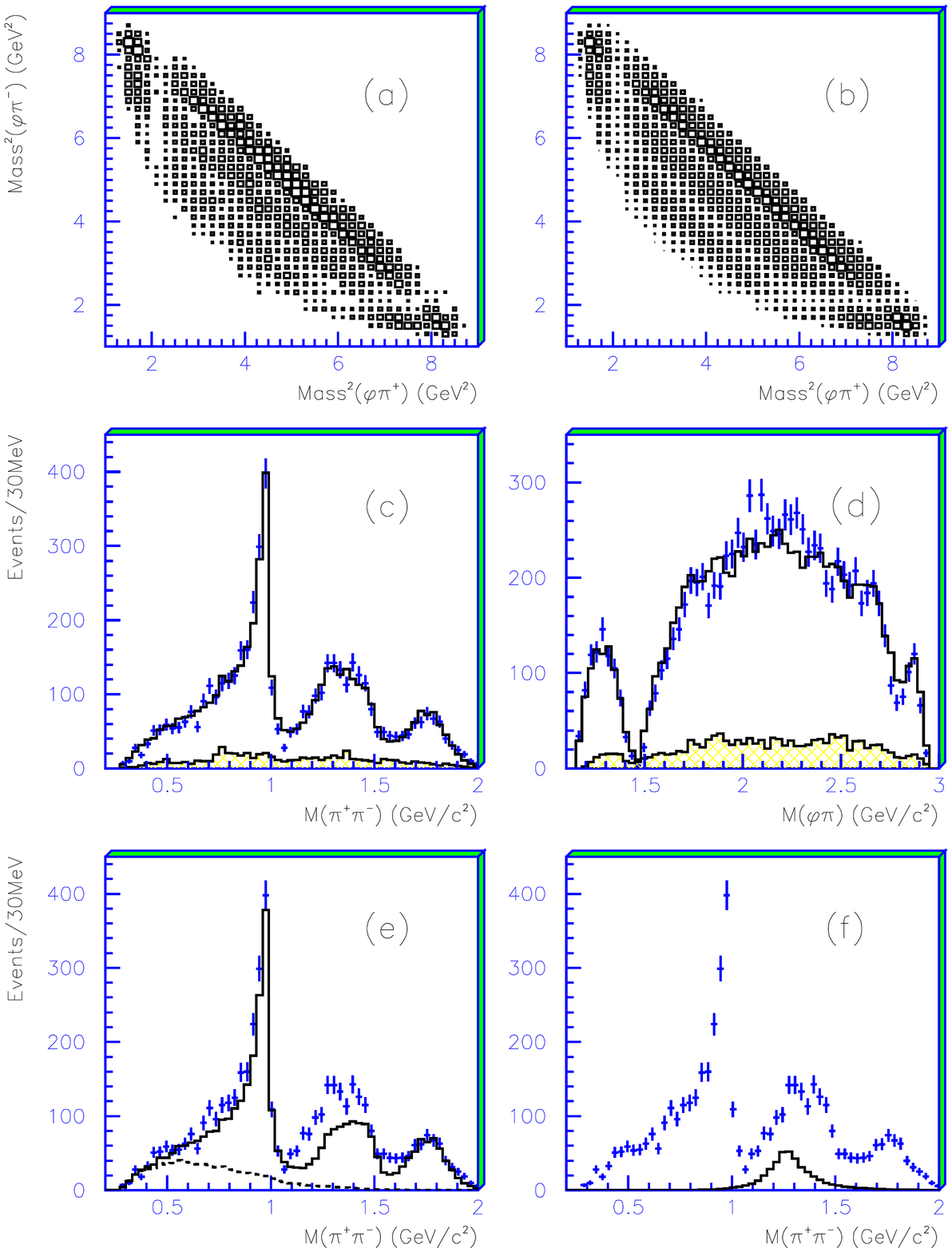,width=11cm}\
\vskip -6mm
\caption{Fit to BES II data on $J/\Psi \to \phi \pi ^+\pi ^-$.
(a) Dalitz plot from data; (b) fitted Dalitz plot;
(c) $\pi \pi$ mass projection and fit (upper histogram);
experimental background is shown by the lower histogram;
(d) fit to the $\phi \pi^+$ mass
projection with the same format as (c);
(e) the coherent sum of $0^+$ contributions (full histogram); the
lower histogram shows the $\sigma$ component;
(f) the contribution from $f_2(1270)$.}
\end {center}
\end {figure}

The fit determines the branching ratio of $\sigma \to \eta \eta$
compared with $\sigma \to \pi ^0 \pi ^0$. There is, however,
a substantial error which arises from interferences between
the three $\pi ^0 \pi ^0$ contributions in $3\pi ^0$ data.
The $\sigma$ contribution is the largest one to
$3\pi ^0$ data, and is subject to some flexibility depending
on the precise $s$-dependence of the amplitude.
There is a subtlety concerning the evaluation of the branching
ratio.
In the $3\pi ^0$ data, the integrated intensity comes not only
from the three individual $\pi \pi $ combinations 12, 23 and 13,
but also from interferences between them. These interferences
are quite large.
What one needs is the branching ratio of an isolated resonance
without these interferences.
Putting this point in a different way, the resonances have
specific coupling constants $g^2$ to each decay channel.
One needs to derive these allowing for the fact that
the three combinations interfere in the integrated intensity.
This is done by evaluating intensities with and without these
interferences.
The result is
\begin {equation}
g^2(\eta\eta)/g^2(\pi \pi) = 0.19 \pm 0.05,
\end {equation}
in close agreement with the earlier determination of Ref. \cite
{Recon}, namely $0.20 \pm 0.05$.

Branching ratios for $f_0(1500)$ and $f_0(1370)$ to $\eta \eta$
compared with $\pi \pi$ are subject to the same large uncertainty
arising from the range of possible fits of $\sigma$ to $3\pi ^0$
data.
Results are
\begin {eqnarray}
\Gamma [f_0(1500) \to \eta\eta)/\Gamma (f_0(1500) \to \pi \pi] &=&
0.135 \pm 0.04, \\
\Gamma [f_0(1370) \to \eta\eta)/\Gamma (f_0(1370) \to \pi \pi] &=&
0.19  \pm 0.07.
\end {eqnarray}

\section {Refitting data on $J/\Psi \to \phi \pi ^+\pi ^-$}
These data provided earlier evidence for $f_0(1370)$ \cite {phipipi}.
They are refitted here using for $\sigma$, $f_0(1370)$ and $f_0(1500)$
the formulae of Section 2 and the parameters determined here.
The fit to $\phi \pi \pi$ data is displayed in Fig. 21, retaining
the layout of the BES publication for comparison.
Data on $\phi KK$ are refitted simultaneously, but changes are
barely visible on figures and will therefore not be shown here.

\begin {figure}  [htb]
\begin {center}
\vskip -24mm
\epsfig{file=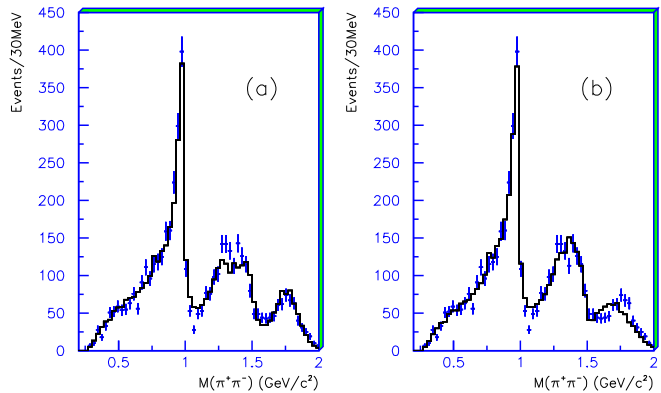,width=9cm}\
\vskip -6mm
\caption{(a) Fit to $\phi \pi \pi$ data without $f_0(1370)$; (b)
the fit with a very broad $0^+$ signal at 1650 MeV interfering
destructively with $f_0(1500)$.}
\end {center}
\end {figure}

Technical details are as follows.
It is important to constrain the branching ratios of $f_2(1270)$ and
$f_0(1500)$ between $KK$ and $\pi \pi$ within one standard deviation
of PDG averages, so as to stabilise the fit to $\phi KK$ data.
The new $\sigma$ parametrisation replaces the previous one, but
has almost no effect: a 0.7 standard deviation change in log
likelihood;
there are much larger changes from reparametrisation of
$f_0(1500)$ and $f_0(1500)$.
The $\sigma \to KK$ contribution is very small and has little effect
on the fit.

A reminder is necessary concerning the $\phi \pi \pi$ data.
The BES analysis located a peak in $\phi \pi$ at 1500 MeV, arising
from the triangle graph due to $J/\Psi \to K^+K^-\pi^+\pi ^-$,
followed by rescattering $KK \to \phi$.
This was eliminated by a kinematic cut,
creating the dips in the $\phi \pi$ mass spectrum of Fig. 21(d)
at 1.5 and 2.8 GeV.

There is a peak at 1350 MeV in the $\pi \pi$ mass projection (c)
attributed to interference between $f_2(1270)$, $f_0(1370)$ and
$f_0(1500)$.
The new fit is shown by the upper histogram; the lower one shows
experimental background.
The fit is marginally poorer than that of the BES publication,
which had the freedom to optimise the mass of $f_0(1370)$ to
$1350 \pm 50$ MeV and the width to $265 \pm 40$ MeV.
It resulted in an $f_0(1370)$ contribution larger than $f_0(1500)$,
though both were needed.
Their roles are now reversed, with a $4.6\%$ intensity contribution
to $\phi \pi ^+\pi ^-$ from $f_0(1370)$ and a 6.1\% contribution
from $f_0(1500)$.
On Fig. 21(e), the full histogram shows the coherent sum of $J^P = 0^+$
contributions and the lower dashed histogram the $\sigma$ contribution.
Fig. 21(f) shows the $f_2(1270)$ contribution.
It is cleanly separated from $0^+$ by angular distributions for
production and decay of resonances.

Fig. 22(a) shows the fit without $f_0(1370)$.
The narrow $f_0(1500)$ is unable to compensate via interferences
with $\sigma$, $f_2(1270)$ and $f_0(980)$ and the
resulting fit is visibly poor.
The fit is worse than Fig. 21 by 34.8 in log likelihood,
i.e. $>8$ standard deviations.
This is sufficient to confirm the presence of $f_0(1370)$, but
not enough to influence its fitted mass and width.

There is evidence that the $f_0(1370)$ is resonant.
If its phase variation is artificially suppressed, leaving the
line-shape unchanged (even though the resulting amplitude is
non-analytic, therefore illogical), the fit is worse by
12.4 in log likelihood, $\sim 5$ standard deviations.

It has been suggested that the $f_0(1790)$ in BES data might
be eliminated by fitting a broad $J^P = 0^+$ $\pi \pi$ signal
with which $f_0(1500)$ interferes destructively, producing an
interference minimum at $\sim 1600$ MeV.
This suggestion was based on the possibility that the high mass
tail of the $\sigma$ might contribute strongly.
That now appears illogical, since the $\sigma$ amplitude falls
rapidly with increasing $s$ and is already small at 1 GeV.

The suggestion has been tested by fitting with a broad component
with a width 1000 MeV and an optimised mass of 1650 MeV.
Fig. 22(b) shows the poor fit.
It is of course possible to tune the width of this broad component
to a lower value and produce a reasonable fit.
However, it makes little sense to invent a new broad component when
there is independent evidence for $f_0(1790)$ with consistent parameters in
$J/\Psi \to \gamma (4\pi)$ data from both Mark III \cite {G4pi1} and
BES I \cite {G4pi2}.
Data on  $J/\Psi \to \gamma \omega \phi$ \cite {omegaphi} exhibit a
striking $0^+$ peak at 1812 MeV, which is consistent within errors with
the upper side of $f_0(1790)$; the $\omega \phi$ threshold is at 1801
MeV.

\section {Other data}
Let us recall that the first evidence for $f_0(1370)$ came from
data at the Argonne and BNL laboratories on $\pi \pi \to KK$ in the
1970 era.
There is a distinct dip between the 1300 MeV peak and the narrow
$f_0(980)$.
Fig. 10(a) of Ref. \cite {Recon} shows that without $f_0(1370)$
this dip cannot be fitted: the $f_0(980)$ is too narrow and
interferences with the broad $\sigma$ and $f_0(1500)$ fail
to fit the data.
Though these data alone at not sufficient proof of the existence of
$f_0(1370)$, they are entirely consistent with parameters fitted
here.
The fit made in Ref. \cite {Recon} has been rerun using the new
parametrisations of the $\sigma$, $f_0(1370)$ and $f_0(1500)$
determined here.
There is no significant change to conclusions of Ref. \cite {Recon}.
The analysis of Crystal Barrel data on $KK\pi$ channels by
Anisovich and Sarantsev \cite {AandS02} finds $f_0(1370)$
mass, width and coupling constant in acceptable agreement with those
reported here.

Data of Barberis et al. \cite {Barberis} on central production of
$\pi \pi$ reveal quite different azimuthal distributions for
$f_0(1370)$ and $f_0(1500)$, suggesting that two separate resonances
are needed, whatever the nature of $f_0(1370)$,

\section {The need for further analyses}
Those who question the existence of $f_0(1370)$ should be
concerned also about the existence of $a_0(1450)$.
Despite its appearance in the Summary Table of the Particle
Data Book, it is subject to the same questions about
dispersive corrections as $f_0(1370)$.
A fresh analysis is needed of data on $\bar pp \to \eta \pi ^0 \pi ^0$,
where it was discovered \cite {a01450}, including $m(s)$ into the
parametrisation. It has also been observed in $\bar pp$ at rest $\to
\omega \rho \pi ^0$ in the $\omega \rho$ channel \cite {a0wr}.
The  threshold for this final state will contribute strongly to $m(s)$,
as may the possible decay channel $a_0(980)\sigma$.
A combined analysis of data on $\eta \eta \pi ^0$ and
$\omega \rho \pi ^0$ is in progress and will be  reported separately.
If the $a_0(1450)$ survives, it is very likely to be a $q\bar q$
state, since none other is available for the isospin 1 component
of the nonet in this mass range.
If so, it is plausible that $f_0(1370)$ is likewise $q\bar q$
(mixed with the glueball).

Next, it is highly desirable to fit all existing data for the
$J^{PC} = 1^{--}$ sector including dispersive corrections.
A start has been made on this type of analysis by Weng et al.
\cite {Weng}.
The greatest need is for data on $\pi \pi \to 4\pi$.
These are needed to pin down details of the $4\pi$ final state
with $J^{PC} = 0^{++}$, $1^{--}$ and $2^{++}$.
Such data were presented in a preliminary form by Ryabchikov
at Hadron95, but have not been the subject of a full publication
yet.
The analysis including full dispersive effects is doubtless a
major undertaking, but even limited information about
the $4\pi$ channel would be very important.

Yet another example which may be affected strongly by dispersive
effects is the $J^{PC} = 1^{-+}$ sector.
There is substantial evidence for $\pi _1 (1600)$
in decays to $b_1(1235)\pi$, $\eta '\pi$, $f_1(1285)\pi$ and
$\rho \pi$.
There is also evidence for structure in $\eta \pi$ at $\sim 1400$
MeV.
This is close to the sharp thresholds for $b_1(1235)\pi$ and
$f_1(1285)\pi$.
It is still an open question whether there really is a
resonant $\pi_1(1400)$, or whether it can be fitted adequately as
a threshold effect. It is also possible that there is a resonance
associated directly with these thresholds.
What is needed is an analysis including dispersive effects due to
the thresholds.

\section {Concluding Remarks}
The $f_0(1370)$ is highly significant statistically in
5 sets of data: $\bar pp \to 3\pi ^0$ at rest in liquid
hydrogen and gas, corresponding data for the $\eta \eta
\pi^0$ data channel, and $J/\Psi \to \phi \pi ^+ \pi ^-$.
Overall, it is statistically more than a 52 standard
deviation effect.
What is also important is that fitted parameters of
$f_0(1370)$ agree remarkably closely between the $3\pi ^0$
data in liquid and gas.
There is weaker but consistent evidence for $f_0(1370)$
in $\pi \pi \to KK$.

The data cannot be explained by the high mass tail of the
$\sigma$, because it is too broad.
The requirement for a peak in data with a full width of 207 MeV
requires an additional narrower state, identified here with
the $f_0(1370)$.
No such pole has appeared within the present
parametrisation of $\sigma \to 4\pi$.

Dispersive contributions due to the opening of the $4\pi$
channel have large effects, renormalising the Breit-Wigner
denominator. This severely limits the range of $\Gamma _{4\pi}/
\Gamma _{2\pi}$ which can be successfully fitted to data.
Despite the strong effect of the $4\pi$ threshold, the resonance
loop on the Argand diagram is very close to a circle.
This provides some justification for the common practice of
fitting a Breit-Wigner resonance of constant width: the
resonance behaves to first approximation like a simple pole.
However, the phase is significantly affected once one reaches
a mass more than one half-width above resonance.
A simple Breit-Wigner amplitude is adequate for finding resonances;
including $m(s)$ uncovers the dynamical effect of the threshold.

There are presently no significant inconsistencies in parameters of
$f_0(1370)$ between sets of data on $\pi \pi$, $KK$ and $\eta \eta$
channels.
Wide variations of mass and width appear only in analyses of
$4\pi$ data.
However, those analyses do not presently include $\sigma \to 4\pi$,
which is found here to play a large role.
Unfortunately, these analyses need to be repeated including the
$\sigma$ contribution.
Disagreements therefore exist only about schemes into which different
authors wish to fit the known states.
Those schemes should not be used as the basis for claiming that
$f_0(1370)$ or any other resonance does not exist.

Some authors, for example Maiani et al. \cite {Maiani}, raise
concerns about mass differences between $f_0(1370)$, $K_0(1430)$
and $a_0(1450)$.
It now appears that $f_0(1370)$ is nearly degenerate in mass
with $a_1(1260)$, $f_1(1285)$ and $f_2(1270)$.
So it appears to pose no particular problem.
It is likely that the mass of $K_0(1430)$ is influenced by its
strong coupling to $K\eta' $, whose threshold opens at $\sim 1450$
MeV.
It is also known that $a_0(1450)$ appears only weakly in the
$\eta \pi$ channel; it is likely that its mass is affected
strongly by coupling to $\omega \rho$ and $a_0(1450)\sigma$ thresholds,
just as the mass of $f_2(1565)$ is close to the $\omega \omega$
threshold, and much lower than the mass of $a_2(1700)$.

The present analysis of elastic data produces new quantitative
estimates of $\Gamma _{2\pi}$ for $f_0(1500)$, $f_2(1565)$,
$\rho_3(1690)$, $\rho_3(1990)$ and $f_4(2040)$.
Data on $\pi \pi \to 4\pi$ would help greatly in confirming
the present analysis, and parametrising more accurately
$1^{--}$ states and $f_2(1565)$.

A final speculative remark emerges from the mixing
between $\sigma$, $f_0(1370) $ and $f_0(1500)$, which appears to
be a necessity in fitting the elastic $\pi \pi$ data.
This mixing suggests a possible analogy with chemical binding.
In the hydrogen molecule, two configurations of protons and electrons
mix.
This is the familiar process of hybridisation.
The lowest eigenstate may be calculated using a Variational Principle.
The suggestion made here is that confinement involves
a hybridisation due to overlapping of nearby resonant states.
One linear combination of states with favourable $SU(3)$
configuration moves down and other combinations are pushed up in
energy, in a way analogous to formation of the covalent chemical bond.
A similar Variational principle is involved in the formation of a
superconductor.

The relevance of such an idea to the confinement process itself
is a matter of conjecture without calculations to support it.
The idea is that there is feedback between the formation
of resonances and a dynamic confinement, i.e. condensation.
Such a mechanism would explain naturally why most of the huge number
of possible molecular states are not observed: most of them are driven
upwards and become a continuum.

\section {Aknowledgements}
It is a pleasure to thank Dr. Andrei Sarantsev for providing the
Crystal Barrel data used here and for discussions about meson resonances
in general over a period of many years.
I am also grateful to Profs. V. Anisovich, G. Rupp and E. van Beveren
for similar discussions.
I wish to thank Prof. E. Klempt for providing
Fig. 2 of unbinned Crystal Barrel data.


\begin {thebibliography}{99}
\bibitem {Glueball} C.J. Morningstar and M.J. Peardon, Phys. Rev.
D {\bf 69} 034509 (1999)                                         
\bibitem{KL1} E. Klempt, hep-ph/0404270                          
\bibitem {KLno1370} E. Klempt and A. Zaitsev, ``Glueballs, hybrids,
multiquarks'', submitted to Phys. Reports (2007)                 
\bibitem {Ochs1} W. Ochs, AIP Conf. Proc. 619, 167 (2002)        
\bibitem {Ochs2} W. Ochs, QCD06, Montpellier, France,
July 3-7, 2006                                                   
\bibitem{Cohen} D. Cohen et al., Phys. Lett. D {\bf 22} 2595 (1980) 
\bibitem{Pawlicki} A.J. Pawlicki et al., Phys. Rev. D {\bf 15}
3196 (1977)                                                        
\bibitem{Etkin} A. Etkin et al., Phys. Rev. D {\bf 25} 1786 (1982) 
\bibitem {Amsler92} C. Amsler et al., Phys. Lett. B {\bf 291} 347
(1992)                                                             
\bibitem {Gaspero} M. Gaspero, Nucl. Phys. A {\bf 562} 407 (1993) 
\bibitem {Obelix1} A. Adamo et al., Nucl. Phys. A {\bf 558} 13c
(1993)                                                            
\bibitem {Meyer} C. Amsler et al., Phys. Lett. B {\bf 322} 431
(1994)                                                            
\bibitem{CBAR1} V.V. Anisovich et al., Phys. Lett. B {\bf 323}
233 (1994)                                                        
\bibitem{CBAR2} V.V. Anisovich, D.V. Bugg, A.V. Sarantsev and B.S. Zou,
Phys. Rev. D {\bf 50} 1972 (1994)                                 
\bibitem{CBAR3} D.V. Bugg, V.V. Anisovich, A.V. Sarantsev and B.S. Zou,
Phys. Rev. D {\bf 50} 4412 (1994)                                 
\bibitem{AmslerD} C. Amsler et al., Physics Letters B {\bf 355}
425 (1995)                                                        
\bibitem {FurtherA} A. Abele et al., Nucl. Phys. A {\bf 609}
562 (1996)                                                        
\bibitem {CM1996} D.V. Bugg, A.V. Sarantsev and B.S. Zou, Nucl. Phys. B
{\bf 471} 59 (1996)                                               
\bibitem {Hyams} B.D. Hyams et al., Nucl. Phys. B {\bf 64} 134 (1973)
\bibitem {PDG} Particle Data Group, J. Phys. G {\bf 33} 1 (2006)
\bibitem{phipipi} M. Ablikim et al., Phys. Lett. B {\bf 607} 243
(2005)                                                            
\bibitem{Recon} D.V. Bugg, Euro. Phys. J C {\bf 47} 45 (2006)     
\bibitem{Nana} A.V. Anisovich et al., Nucl. Phys. A {\bf 690}
567 (2001)                                                        
\bibitem{E791} E.M. Aitala et al., Phys. Rev. Lett. {\bf 86} 770
(2001)                                                            
\bibitem{ompipi} M. Ablikim et al., Phys. Lett. B {\bf 598} 149
(2004)                                                            
\bibitem{sigphase} D.V. Bugg, Euro. Phys. J C {\bf 37} 433 (2004) 
\bibitem{TornqvistA} N.A. Tornqvist, Phys. Rev. Lett. {\bf 49}    
624 (1982)
\bibitem{TornqvistB} N.A. Tornqvist, Z. Phys. C {\bf 68}  647 (1995) 
\bibitem{Maiani} L. Maiani et al., Eur. Phys. J C {\bf 50} 609
(2007)                                                            
\bibitem{AandS02} V.V. Anisovich and A.V. Sarantsev,
Eur. Phys. J A {\bf 16} 229 (2003) and further references given there
\bibitem{WKK} M. Ablikim et al., Phys. Lett. B {\bf 603} 138
(2004)                                                            

\bibitem{sigpole} D.V. Bugg, J. Phys. G {\bf 34} 151 (2007)       
\bibitem{AAS97} A.V. Anisovich, V.V. Anisovich and A.V. Sarantsev,
Zeit. Phys. A {\bf 359} 173 (1997)                                
\bibitem{Hodd} C.A. Baker et al., Physics Letters B {\bf 467}
147 (1999)                                                        
\bibitem {Stark} G. Reifenrother and E. Klempt, Nucl. Phys. A {\bf 503}
886 (1989)                                                        
\bibitem {Asterix} B. May et al., Phys. Lett.  B {\bf 225}
450 (1989)                                                        
\bibitem {Ishida} M. Ishida et al., Prog, Theor. Phys {\bf 104}
203 (2000)                                                        
\bibitem{Pislak} S. Pislak et al., Phys. Rev. D {\bf 67}
072004 (2003)                                                     
\bibitem{Caprini} I. Caprini, I. Colangelo and H. Leutwyler, Phys. Rev.
Lett. {\bf 96} 032001 (2006)                                      
\bibitem{WuZou} F.Q. Wu et al., Nucl. Phys. A {\bf 735} 111 (2004) 
\bibitem{Pelaez} J.R. Pelaez, hep-ph/0510215                       
\bibitem{BABAR} B. Aubert et al., Phys. Rev. D {\bf 71}
052001 (2005)                                                      
\bibitem{Babrho} B. Aubert et al., Phys. Rev. D {\bf 73} 052003
(2006)                                                             
\bibitem{Frabetti} P.L. Frabetti et al., Phys. Lett. B {\bf 578}
290 (2004)                                                         
\bibitem{G4pi1} D.V. Bugg et al., Phys. Lett. B {\bf 353}
378 (1995)                                                         
\bibitem{G4pi2} J.Z. Bai et al., Phys. Lett. B {\bf 472}
207 (1999)                                                         
\bibitem{omegaphi} M. Ablikim et al., Phys. Rev. Lett. 96 162002
(2006)                                                             
\bibitem{Barberis} D. Barberis et al., Phys. Lett. B {\bf 474}
423 (2000)                                                         
\bibitem{a01450} C. Amsler et al., Phys. Lett. B {\bf 333}
277 (1994)                                                         
\bibitem{a0wr} C.A. Baker et al.,  Phys. Lett. B 563
140 (2003)                                                         
\bibitem{Weng} Y. Weng et al., hep-ex/0512052                      
\end {thebibliography}
\end {document}